\documentclass[11pt,onecolumn,draftclsnofoot]{IEEEtran}
\usepackage{graphicx}
\usepackage{epsfig}
\usepackage{subfigure}
\usepackage{amssymb}
\usepackage{amsbsy}
\usepackage{amsmath}
\usepackage{cite}
\usepackage{stfloats}

\newcommand{\E}{\ensuremath{\hbox{\textbf{E}}}}

\newtheorem{theorem}{Theorem}
\newtheorem{lemma}{Lemma}

\newtheorem{definition}{Definition}

\newcommand{\beq}{\begin{equation}}
\newcommand{\eeq}{\end{equation}}
\newcommand{\bea}{\begin{array}}
\newcommand{\ena}{\end{array}}
\newcommand{\bds}{\begin {itemize}}
\newcommand{\eds}{\end {itemize}}
\newcommand{\bdf}{\begin{definition}}
\newcommand{\blm}{\begin{lemma}}
\newcommand{\edf}{\end{definition}}
\newcommand{\elm}{\end{lemma}}
\newcommand{\bthm}{\begin{theorem}}
\newcommand{\ethm}{\end{theorem}}
\newcommand{\bprp}{\begin{prop}}
\newcommand{\eprp}{\end{prop}}
\newcommand{\bcl}{\begin{claim}}
\newcommand{\ecl}{\end{claim}}
\newcommand{\bcr}{\begin{coro}}
\newcommand{\ecr}{\end{coro}}
\newcommand{\bquest}{\begin{question}}
\newcommand{\equest}{\end{question}}

\newcommand{\larrow}{{\larrow}}

\newcommand{\nin}{{\not \in}}

\def\urltilda{\kern -.15em\lower .7ex\hbox{\~{}}\kern .04em}

\begin{document}\title{Optimal Index Policies for Anomaly Localization in Resource-Constrained Cyber Systems}
\author{Kobi Cohen$^1$, Qing Zhao$^1$, \emph{Fellow}, IEEE, and Ananthram Swami$^2$, \emph{Fellow}, IEEE
\thanks{This work was supported by Army Research Laboratory under Grant W911NF1120086 and by the National Science Foundation under Grant CCF-1320065.}
\thanks{Part of this work was presented at the IEEE Global Conference on Signal and Information Processing (GlobalSIP), Austin, Texas, USA, December 2013.}
\thanks{$^1$ Department of Electrical and Computer Engineering, University of California, Davis, CA 95616 USA (e-mail: yscohen@ucdavis.edu; qzhao@ucdavis.edu).}
\thanks{$^2$ Army Research Laboratory, Adelphi, MD 20783 USA (e-mail: a.swami@ieee.org).}
}
\date{}
\maketitle
%
%-------------abstract----------------------------------
%-------------------------------------------------------
\begin{abstract}
\label{sec:abstract}
The problem of anomaly localization in a resource-constrained cyber system is considered. Each anomalous component of the system incurs a cost per unit time until its anomaly is identified and fixed. Different anomalous components may incur different costs depending on their criticality to the system. Due to resource constraints, only one component can be probed at each given time. The observations from a probed component are realizations drawn from two different distributions depending on whether the component is normal or anomalous. The objective is a probing strategy that minimizes the total expected cost, incurred by all the components during the detection process, under reliability constraints. We consider both independent and exclusive models. In the former, each component can be abnormal with a certain probability independent of other components. In the latter, one and only one component is abnormal. We develop optimal index policies under both models. The proposed index policies apply to a more general case where a subset (more than one) of the components can be probed simultaneously. The problem under study also finds applications in spectrum scanning in cognitive radio networks and event detection in sensor networks.
\end{abstract}
%-------------end abstract------------------------------
%
% Keyword section
\def\keywords{\vspace{.5em}
{\bfseries\textit{Index Terms}---\,\relax%
}}
\def\endkeywords{\par}
\keywords
Anomaly localization, sequential hypothesis testing, Sequential Probability Ratio Test (SPRT), composite hypothesis testing.
\section{Introduction}
\label{sec:intro}

Consider a cyber system with $K$ components. Each component may be in a normal or an abnormal state. If abnormal, component $k$ incurs a cost $c_k$ per unit time until its anomaly is identified and fixed. Due to resource constraints, only one component can be probed at a time, and switching to a different component is allowed only when the state of the currently probed component is declared. The observations from a probed component (say $k$) follow distributions $f_k^{(0)}$ or $f_k^{(1)}$ depending on whether the component is normal or anomalous, respectively. The objective is a probing strategy that dynamically determines the order of the sequential tests performed on all the components so that the total cost incurred to the system during the entire detection process is minimized under reliability constraints.

\subsection{Main Results}

The above problem presents an interesting twist to the classic sequential hypothesis testing problem. In the case when there is only one component, minimizing the cost is equivalent to minimizing the detection delay, and the problem is reduced to a classic sequential test where both the simple and the composite hypothesis cases have been well studied. With multiple components, however, minimizing the detection delay of each component is no longer sufficient. The key to minimizing the total cost is the order at which the components are being tested. It is intuitive that we should prioritize components that incur higher costs when abnormal as well as components with higher prior probabilities of being abnormal. Another parameter that plays a role in the total system cost is the expected time in detecting the state of a component, which depends on the observation distributions $\{f_k^{(0)},\, f_k^{(1)}\}$. It is desirable to place components that require longer testing time toward the end of the testing process. The challenge here is how to balance these parameters in the dynamic probing strategy.

We show in this paper that the optimal probing strategy is an open-loop policy where the testing order can be predetermined, independent of the realizations of each individual test in terms of both the test outcome and the detection time.
Furthermore, the probing order is given by a simple index. Specifically, under the independent model where each component is abnormal with a prior probability $\pi_k$ independent of other components, the index is in the form of $\pi_k c_k/\mathbf{E}(N_k)$, where $\mathbf{E}(N_k)$ is the expected detection time for component $k$. Under the exclusive model where one and only one component is abnormal, the index is in the form of $\pi_k c_k/\mathbf{E}(N_k|H_0)$ where $\mathbf{E}(N_k|H_0)$ is the expected detection time for component $k$ under the hypothesis of it being normal. These index forms give a clean expression on how the three key parameters---the cost, the prior probability, and the difficulty in distinguish normal distribution $f_k^{(0)}$ from abnormal distribution $f_k^{(1)}$---are balanced in choosing the probing order. Furthermore, it is interesting to notice the difference in the indices for these two models. Intuitively speaking, under the independent model, the detection time of any component, normal or abnormal, adds to the delay in catching the next abnormal component, while under the exclusive model, only the detection times of components in a normal state adds to the delay in catching the abnormal component.

The above simple index forms of the probing order are optimal for both the simple hypothesis ($\{f_k^{(0)},\, f_k^{(1)}\}_{k=1}^K$ are known) and the composite hypothesis ($\{f_k^{(0)},\, f_k^{(1)}\}_{k=1}^K$ have unknown parameters) cases. These index policies also apply to the case where multiple components can be probed simultaneously. While the optimality of these indices in this case remains open, simulation examples demonstrate their strong performance.

\subsection{Applications}
\label{sec:intro_app}

The problem considered here finds applications in anomaly detection in cyber systems, spectrum scanning in cognitive radio systems, and event detection in sensor networks. In the following we give two specific examples.

Consider a cyber network consisting of $K$ components (which can be routers, paths, etc.). Due to resource constraints, only a subset of the components can be probed at a time. An Intrusion Detection System (IDS) analyzes the traffic over the components to detect Denial of Service (DoS) attacks (such attacks aim to overwhelm the component with useless traffic to make it unavailable for its intended use). Let the cost $c_k$ be the normal expected traffic (packets per unit time) over component $k$. Thus, in this example minimizing the total expected cost minimizes the total expected number of failed packets in the network during DoS attacks. The exclusive model applies to cases where an intrusion to a subnet, consisting of $K$ components, has been detected and the probability of each component being compromised is small (thus with high probability, there is only one abnormal component).

Another example is spectrum sensing in cognitive radio systems. Consider a spectrum consisting of $K$ orthogonal channels. Accessing an idle channel leads to a successful transmission, while accessing a busy channel results in a collision with other users. A Cognitive Radio (CR) is an intelligent device that can detect and access idle channels in the wireless spectrum \cite{Zhao_2007_Survey}. Due to resource constraints, only a subset of the channels can be sensed at a time. Once a channel is identified as idle, the CR transmits over it. Let $c_k$ be the achievable rate over channel $k$. Thus, in this example minimizing the total expected cost minimizes the total expected loss in data rate during the spectrum sensing process.

\subsection{Related Work}
\label{ssec:related}

The classic sequential hypothesis testing problem which pioneered by Wald \cite{Wald_1947_Sequential} considers only a single stochastic process. For simple binary hypothesis testing, Wald showed that the Sequential Probability Ratio Test (SPRT) is optimal in terms of minimizing the expected sample size under given type $I$ and type $II$ error probability constraints. Various extensions for M-ary hypothesis testing and composite hypothesis testing were studied in \cite{Schwarz_1962_Asymptotic, Robbins_1974_Expected, Lai_1988_Nearly, Pavlov_1990_Sequential, Lai_1994_Nearly, Tartakovsky_2002_Efficient, Draglin_1999_Multihypothesis} for a single process. In these cases, asymptotically optimal performance can be obtained as the error probability approaches zero.

A number of studies exist in the literature that consider sequential detection over multiple processes. Differing from this work that focuses on minimizing the total cost incurred by anomalous components, these existing results adopt the objective of minimizing the total detection delay. In particular, the problem of quickly detecting an idle period over multiple independent ON/OFF processes was considered in \cite{Zhao_2010_Quickest} where a threshold policy was shown to be optimal. The ON/OFF nature of the processes and the objective of minimizing the total detection delay make the problems considered in \cite{Zhao_2010_Quickest} fundamentally different from the one considered in this work. In \cite{Li_2009_Restless}, the problem of quickest detection of the emergence of primary users in multi-channel cognitive radio networks was considered. In \cite{Caromi_2013_Fast}, the problem of quickest detection of idle channels over $K$ independent channels was studied. The idle/busy state of each channel was assumed fixed over time, and the objective was to minimize the detection delay under error constraints. It was shown that the optimal policy is to carry out an independent SPRT over each channel, irrespective of the testing order. In contrast to \cite{Caromi_2013_Fast}, we show in this paper that the optimal policy in our model highly depends on the testing order even when the processes are independent. In \cite{Lai_2011_Quickest}, the problem of identifying the first abnormal sequence among an infinite number of i.i.d sequences was considered. An optimal cumulative sum (CUSUM) test was established under this setting. Variations of the latter model have been studied in \cite{Malloy_2012_Quickest, Tajer_2013_Quick}. The sequential search problem under the exclusive model was investigated in \cite{Zigangirov_1966_Problem, Klimko_1975_Optimal, Dragalin_1996_Simple, Stone_1971_Optimal}. Optimal policies were derived for the problem of quickest search over Weiner processes \cite{Zigangirov_1966_Problem, Klimko_1975_Optimal, Dragalin_1996_Simple}. It was shown in \cite{Zigangirov_1966_Problem, Klimko_1975_Optimal} that the optimal policy is to select the sequence with the highest posterior probability of being the target at each given time. In \cite{Dragalin_1996_Simple}, an SPRT-based solution was derived, which is equivalent to the optimal policy in the case of searching over Weiner processes. However, minimizing the total expected cost in our model leads to a different problem and consequently a different index policy.

The classic target whereabouts problem is also a detection problem over multiple processes. In this problem, multiple locations are searched to locate a target. The problem is often considered under the setting of fixed sample size as in \cite{Tognetti_1968_An, Kadane_1971_Optimal, Castanon_1995_Optimal, Zhai_2013_Dynamic}. In \cite{Tognetti_1968_An, Kadane_1971_Optimal, Zhai_2013_Dynamic}, searching in a specific location provides a binary-valued measurement regarding the presence or absence of the target. In \cite{Castanon_1995_Optimal}, Castanon considered the dynamic search problem under continuous observations: the observations from a location without the target and with the target have distributions $f$ and $g$, respectively. The optimal policy was established under a symmetry assumption that $f(x)=g(b-x)$ for some $b$.

The anomaly detection problem may also be considered as a variation of active hypothesis testing in which the decision maker chooses and dynamically changes its observation model among a set of observation options. Classic and more recent studies of general active hypothesis testing problems can be found in
\cite{Blackwell_1953_Equivalent, Chernoff_1959_Sequential, DeGroot_1962_Uncertainty, Naghshvar_2013_Active, Nitinawarat_2013_Controlled, Cohen_2014_Active, Cohen_2014_Quickest}.

\subsection{Organization}
\label{ssec:organization}

In Section \ref{sec:network} we describe the system model and problem formulation. In Section \ref{sec:two_stage} we propose a two-stage optimization problem that simplifies computation while preserving optimality. In Section \ref{sec:simple} we derive optimal algorithms under the independent and exclusive models for the simple hypothesis case. In Section \ref{sec:uncertainty} we extend our results to the composite hypothesis case: we derive asymptotically optimal algorithms under the independent and exclusive models. In Section \ref{sec:simulation} we provide numerical examples to illustrate the performance of the algorithms.

\section{System Model and Problem Formulation}
\label{sec:network}
Consider a cyber system consisting of $K$ components, where each component may be in a normal state or abnormal state. Define
\beq
\bea{l}
\mathcal{H}_0\triangleq\left\{k: 1\leq k \leq K \; , \; \mbox{component $k$ is healthy}
\right\} \;, \vspace{0.1cm} \\
\mathcal{H}_1\triangleq\left\{k: 1\leq k \leq K \; , \; \mbox{component $k$ is abnormal}
 \right\} \;,
\ena
\eeq
as the sets of the normal and abnormal components.\\
We consider two different anomaly models.
\begin{enumerate}
  \item Exclusive model: One and only one component is abnormal; the \emph{a priori} probability that component $k$ is the abnormal one is $\pi_k$, where $\sum_{k=1}^{K}{\pi_k}=1$. \vspace{0.0cm}
  \item Independent model: Each component $k$ is abnormal with \emph{a priori} probability $\pi_k$ independent of other components. \vspace{0.0cm}
\end{enumerate}
Every abnormal component $k$ incurs a cost $c_k$ ($0\leq c_k<\infty$) per unit time until it is tested and identified. Components in a normal state do not incur cost. We focus on the case where only one component can be probed at a time. The resulting probing strategies apply to the case where a subset of the components can be probed simultaneously and their performance in this case are studied via simulation examples, given in Section \ref{sec:simulation}. When component $k$ is tested at time $t$, a measurement (or a vector of measurements) $y_k(t)$ is obtained and is independently over time. If component $k$ is healthy, $y_k(t)$ follows distribution $f_k^{(0)}$; if component $k$ is abnormal, $y_k(t)$ follows distribution $f_k^{(1)}$. We focus first on the simple hypothesis case, where the distributions $f_k^{(0)}$, $f_k^{(1)}$ are known. In Section \ref{sec:uncertainty} we extend our results to the composite hypothesis case, where the distributions have unknown parameters. We consider the case where switching across components is allowed only when the state of the currently probed component is declared.

The detection process starts at time $t=1$. The random sample size required to make a decision regarding the state of component $k$ is denoted by $N_k$. We define $\tau_k$ as the stopping time (counted from the beginning of the first test at $t=1$), at which the decision maker stops taking observations from component $k$ and declares its state. The vector of stopping times for the $K$ components is denoted by $\boldsymbol\tau=(\tau_1, ..., \tau_K)$. For example, assume that $K=3$ and the decision maker tests the components according to the following order: $3, 1, 2$. Then, $\tau_3=N_3$, $\tau_1=N_3+N_1$, $\tau_2=N_3+N_1+N_2$.

Let $\delta_k\in\left\{0, 1\right\}$ be a decision rule, which the decision maker uses to declare the state of component $k$ at time $\tau_k$. $\delta_k=0$ if the decision maker declares that component $k$ is in a healthy state (i.e., $H_0$), and $\delta_k=1$ if the decision maker declares that component $k$ is in an abnormal state (i.e., $H_1$).
The vector of decision rules for the $K$ components is denoted by $\boldsymbol\delta=(\delta_1, ..., \delta_K)$.

Let $\mathcal{K}(t)$ be the set of components whose states have not been declared by time $t$ and $\phi(t)$ the index of the component being tested at time $t$ (i.e., a selection rule). Let $\mathbf{y}(t)=\left\{\phi(i), y_{\phi(i)}(i)\right\}_{i=1}^t$ be the set of all observations and actions up to time $t$. The selection rule $\phi(t)$ is a mapping from $\mathbf{y}(t-1)$ to $\mathcal{K}(t)$, indicating which component is chosen to be tested at time $t$ among the components whose states have not been determined. Since switching across components is allowed only when the state of the currently probed component is declared, the selection rule satisfies $\phi(\tau_k-t)=\phi(\tau_k)$ for all $1\leq t\leq N_{\phi(\tau_k)}-1$, $k=1, 2, ..., K$. The vector of selection rules over the time series is denoted by $\boldsymbol\phi=(\phi(1), \phi(2)...)$. An admissible strategy $\mathbf{s}$ is a sequence of $K$ sequential tests for the $K$ components and denoted by the tuple $\mathbf{s}=\left(\boldsymbol{\tau}, \boldsymbol{\delta}, \boldsymbol{\phi}\right)$.

The problem is to find a strategy $\mathbf{s}$ that minimizes the total expected cost, incurred by all the abnormal components during the entire detection process, subject to type $I$ (false-alarm) and type $II$ (miss-detect) error constraints for each component:
\beq\label{eq:opt1}
\bea{lll}
\displaystyle\inf_{\mathbf{s}} & \E\left\{\displaystyle\sum_{k\in \mathcal{H}_1}{c_k\tau_k}\right\} & \vspace{0.1cm} \\
s.t.           &  P_k^{FA}\leq \alpha_k \;\;,\;\;P_k^{MD}\leq \beta_k  & \forall k=1, ..., K  \;,
\ena
\eeq
We point out that the total cost defined in (\ref{eq:opt1}) does not include the cost incurred by miss-detected abnormal components. Since the error constraints are typically required to be small, (\ref{eq:opt1}) well approximates the actual loss in practice.

We have adopted a model where switching across components is allowed only when the test of a currently chosen component is complete. This model is desirable in practical scenarios when switching among components results in additional cost or delay. This model also reduces the memory requirement since only observations from a single component need to be stored. Furthermore, this model is advantageous from a computational complexity perspective. Detection problems involving multiple processes are partially-observed Markov decision processes (POMDP) \cite{Castanon_1995_Optimal} which have exponential complexity in general. As a result, computing optimal policies is intractable (except for some special observation distributions as considered in \cite{Zigangirov_1966_Problem, Castanon_1995_Optimal}). Thus, simplifying the search model is necessary to make the problem mathematically tractable and provide insights and general design guidelines. Similar assumptions have been adopted in \cite{Stone_1971_Optimal, Dragalin_1996_Simple, Lai_2011_Quickest} to simplify the search model under different objectives.

\section{Decoupling of Ordering and Sequential Testing}
\label{sec:two_stage}

In this section, we show that the probing order and the sequential testing of each component can be decoupled. As a consequence, the solution to (\ref{eq:opt1}) can be obtained in two stages.

In the first stage, we solve the following optimization problem for every component $k$:
\beq\label{eq:sub_k_opt}
\bea{l}
\displaystyle\displaystyle\inf_{N_k, \delta_k}{\mathbf{E}(N_k|H_i)} \;, \;\;i=0, 1  \vspace{0.1cm} \\
s.t. \hspace{0.5cm}
P_k^{FA}\leq \alpha_k \;\;,\;\; P_k^{MD}\leq \beta_k    \;.\\ \hspace{1cm}
\ena
\eeq

In the second stage, the problem is to find a selection rule $\boldsymbol{\phi}$ that minimizes the objective function, given the solution to the $K$ subproblems specified in (\ref{eq:sub_k_opt}):
\beq\label{eq:sub_opt}
\bea{l}
\displaystyle\inf_{\boldsymbol{\phi}} \;\; \E\left\{\displaystyle\sum_{k\in \mathcal{H}_1}{c_k\tau_k\;\left|\;\left(\boldsymbol{N}^*, \boldsymbol{\delta}^*\right)\right.}\right\}
\ena
\eeq
where
\beq
\label{eq:stopping decision_stars}
\boldsymbol{N}^*=(N_1^*, ..., N_K^*) \;,\; \boldsymbol{\delta}^*=(\delta_1^*, ..., \delta_K^*)
\eeq
denote the vectors of stopping times and decision rules, respectively, that solve the $K$ subproblems given in (\ref{eq:sub_k_opt}). Note that the stopping times $\boldsymbol{\tau}=(\tau_1, ..., \tau_K)$ are completely specified by $\boldsymbol{N}^*$ and the selection rule $\boldsymbol{\phi}^*$ that solves (\ref{eq:sub_opt}).

The formulation of the two-stage optimization problem allows us to decompose the original optimization problem (\ref{eq:opt1}) into $K+1$ subproblems (\ref{eq:sub_k_opt}) and (\ref{eq:sub_opt}). In subsequent sections we show that the two-stage optimization problem preserves optimality under both the independent and exclusive models.

\section{The Simple Hypothesis Case}
\label{sec:simple}

In this section we derive optimal solutions to both the independent and exclusive models when the observation distributions under both hypotheses are completely known. We discuss the implementation of the optimal policies in Section \ref{ssec:computing}.

\subsection{SPRT for Each Component}
\label{ssec:SPRT}

For the simple hypothesis case, the solution to the first stage optimization problem (\ref{eq:sub_k_opt}) is given by the SPRT \cite{Wald_1947_Sequential} as follows. \\
Assume that the state of component $j$ has been declared at time $\tau_j$ and component $k$ is chosen to be tested at time $\tau_j+1$. Let
\beq
\label{eq:LR}
L_k(n)=\displaystyle\frac{\prod_{t=\tau_j+1}^{\tau_j+n}{f_k^{(1)}(y_k(t))}}
                           {\prod_{t=\tau_j+1}^{\tau_j+n}{f_k^{(0)}(y_k(t))}}
\eeq
be the Likelihood Ratio (LR) between the two hypotheses for component $k$ at stage $n$.

In SPRT, the stopping and decision rules are given by comparing the LR with boundary values at each stage $n$ \cite{Wald_1947_Sequential}. Specifically, let $A_k, B_k$ ($B_k>1/A_k$) be the boundary values used by the SPRT for component $k$. The SPRT algorithm is carried out as follows:
\begin{itemize}
  \item If $L_k(n)\in\left((A_k)^{-1}, B_k\right)$, continue to take observations from component $k$.
  \item If $L_k(n)\geq B_k$, stop taking observations from component $k$ and declare it as abnormal (i.e., $\delta_k=1$). Clearly, $N_k=n$.
  \item If $L_k(n)\leq (A_k)^{-1}$, stop taking observations from component $k$ and declare it as normal (i.e., $\delta_k=0$). Clearly, $N_k=n$. %\vspace{0.2cm}
\end{itemize}

Implementation of the SPRT requires the computation of $A_k$ and $B_k$ to ensure the constraints on the error probabilities. In general, the exact determination of the boundary values is laborious and depends on the observation distribution. Wald's approximation can be applied to simplify the computation \cite{Wald_1947_Sequential}:
\beq
\label{eq:boundary_approx}
\bea{l}
B_k\approx \displaystyle\frac{1-\beta_k}{\alpha_k} \;\;\;,\;\;\;
A_k\approx \displaystyle\frac{1-\alpha_k}{\beta_k} \;.
\ena
\eeq
Wald's approximation performs well for small $\alpha_k, \beta_k$ and is asymptotically optimal as $\alpha_k, \beta_k$ approach zero. Since type $I$ and type $II$ errors are typically required to be small, Wald's approximation is widely used in practice \cite{Wald_1947_Sequential}.

\subsection{Optimal Index Policies}
\label{ssec:optimal}

We now consider the second stage optimization problem specified by (\ref{eq:sub_opt}) and (\ref{eq:stopping decision_stars}). Our main result is to establish the optimal selection rule as the $\pi c N$-rule for the independent model and the $\pi c N_0$ rule for the exclusive model. Specifically, the $\pi c N$-rule dictates that the components be tested in a decreasing order of $\pi_k c_k/\mathbf{E}(N_k)$ and the $\pi c N_0$-rule dictates that the components be tested in a decreasing order of $\pi_k c_k/\mathbf{E}(N_k|H_0)$. Note that these optimal selection rules are open loop policies: the testing orders can be determined offline (see Section \ref{ssec:computing} for the computation of the indices). With the optimal solution to (\ref{eq:sub_k_opt}), the optimal anomaly detection strategy is to carry out a series of SPRTs on the components ordered according to either the $\pi c N$-rule or the $\pi c N_0$-rule. The resulting strategies are thus referred to as $\pi c N$-SPRT and $\pi c N_0$-SPRT.

The index selection rules $\pi c N$ and $\pi c N_0$ are intuitively satisfying. The priority of component $k$ in terms of testing order should be higher as the cost $c_k$ increases, or the \emph{a priori} probability of being abnormal $\pi_k$ increases. Under the independent model, the priority of component $k$ in terms of testing order should be higher as the expected sample size $\mathbf{E}(N_k)$ decreases (since $\mathbf{E}(N_k)$ contributes to the cost of every component which is tested after component $k$). On the other hand, under the exclusive model, the priority of component $k$ in terms of testing order depends on $\mathbf{E}(N_k|H_0)$ rather than $\mathbf{E}(N_k)$. The reason is that if component $k$ is abnormal, there is no additional cost, incurred by other components (since only one component is abnormal). On the other hand, if component $k$ is healthy, then $\mathbf{E}(N_k|H_0)$ contributes to the cost of the components (which may be abnormal) tested after component $k$.

The optimality of the algorithms is shown in the following theorem.

\textsl{\theorem\label{th:optimality_alg1_2}{
Under the independent and exclusive models, the $\pi c N$-SPRT and $\pi c N_0$-SPRT algorithms, respectively, solve the original optimization problem (\ref{eq:opt1}).
}}\vspace{0.1cm} \\
\begin{proof}
See Appendices \ref{app:exclusive} and \ref{app:independent}. \vspace{0.1cm}
\vspace{0.1cm}
\end{proof}

While $\pi c N$-rule and $\pi c N_0$-rule are open-loop policies, Theorem \ref{th:optimality_alg1_2} shows that they are optimal among the class of both open-loop and closed-loop selection rules. It should be noted that open-loop policies may not preserve optimality under non-linear cost functions or other correlated models. In these cases, the optimal testing order might need to be updated dynamically based on the realizations of each individual test in terms of the test outcome or the detection time.

The $\pi c N$-rule and $\pi c N_0$-rule bear some similarity with the result developed in \cite{Smith_1956_Various}. In \cite{Smith_1956_Various}, the problem of ordering independent operations with given processing times was considered. It was shown that the optimal selection rule for the problem of minimizing an expected weighted sum of completion times of all the operations is to select the components in decreasing order of $c_k/\mathbf{E}(N_k)$, where $c_k$ and $\mathbf{E}(N_k)$ are the weight and the expected processing time for operation $k$, respectively. However, the problem in (\ref{eq:sub_opt}) is different. First, each component may be normal or abnormal (rather than a given processing time with a fixed distribution) and the expected sample size depends on the component state. Second, the objective is to minimize an expected weighted sum of stopping times of abnormal components only. Third, under the exclusive model, the states of the $K$ components are \emph{dependent}. Furthermore, the original optimization (\ref{eq:opt1}) also includes the stopping rules which control the expected sample size.

\subsection{Computing the Indices}
\label{ssec:computing}

Arranging the components according to $\pi c N$-rule or $\pi c N_0$-rule can be done in $O(K\log K)$ time via sorting algorithms. However, computing the expected sample size $\mathbf{E}(N_k|H_i)$ for all $k=1, 2, ..., K$ can be involved. In general, it is difficult to obtain a closed-form expression for $\mathbf{E}(N_k|H_i)$. One way to evaluate $\mathbf{E}(N_k|H_i)$ is to perform off-line simulations (i.e., carrying out $K$ independent SPRTs for the $K$ components). Another way to evaluate $\mathbf{E}(N_k|H_i)$ is to use a closed-form approximation as follows. Since the solution to (\ref{eq:sub_k_opt}) is given by the SPRT, Wald's approximation can be applied \cite{Wald_1947_Sequential}. For every $i ,j =0 ,1$, let
\beq
\label{eq:KL}
D_k(i||j)=\mathbf{E}_i\left(\log\frac{f_k^{(i)}(y_k(1))}{f_k^{(j)}(y_k(1))}\right)
\eeq
be the Kullback-Leibler (KL) divergence between the hypotheses $H_i$ and $H_j$, where the expectation is taken with respect to $f_k^{(i)}$.\\
The expected sample size conditioned on each hypothesis is well approximated by \cite{Wald_1947_Sequential}:
\beq
\label{eq:sample_size_approx_H}
\bea{l}
\mathbf{E}(N_k|H_0)\approx \displaystyle\frac{\left(1-\alpha_k\right)\log \tilde{A}_k - \alpha_k \log \tilde{B}_k}{D_k(0||1)} \;, \vspace{0.1cm} \\
\mathbf{E}(N_k|H_1)\approx \displaystyle\frac{\left(1-\beta_k\right)\log \tilde{B}_k - \beta_k \log \tilde{A}_k}{D_k(1||0)} \;,
\ena
\eeq
where $\tilde{A}_k=(1-\alpha_k)/\beta_k, \tilde{B}_k=(1-\beta_k)/\alpha_k$ are the approximations to $A_k, B_k$, given in (\ref{eq:boundary_approx}). Note that (\ref{eq:sample_size_approx_H}) approach the exact expected sample sizes $\mathbf{E}(N_k|H_0)\rightarrow-\log\beta_k/D_k(0||1)$, $\mathbf{E}(N_k|H_1)\rightarrow-\log\alpha_k/D_k(1||0)$ as the error constraints approach zero. \\
The expected sample size required to make a decision regarding the state of component $k$ is given by:
\beq
\label{eq:sample_size_approx}
\bea{l}
\mathbf{E}(N_k) %\vspace{0.3cm} \\ \hspace{0.5cm}
= \pi_k\mathbf{E}(N_k|H_1)+(1-\pi_k)\mathbf{E}(N_k|H_0) \;,
\ena
\eeq
where the approximation approaches the exact expected sample size for small $\alpha_k, \beta_k$.

Note that optimality of the algorithms is preserved as long as the \emph{order} of the indices is preserved (i.e., the exact index values are not required for optimality). Therefore, optimality can be achieved in practice even when Wald's approximation is used.

\section{The Composite Hypothesis Case}
\label{sec:uncertainty}

In the previous section we focused on the simple hypothesis case, where the distribution under both hypotheses are completely known. For this case, the SPRT was applied to solve (\ref{eq:sub_k_opt}). However, in numerous cases there is uncertainty in the observation distributions.

For example, Consider a one-parameter distribution $f\left(y|\theta_k\right)$, where it is required to test $\theta_k<\theta_k^{(0)}$ against $\theta_k>\theta_k^{(1)}>\theta_k^{(0)}$. As discussed in \cite{Wald_1947_Sequential}, the SPRT can be applied to this problem by testing $\theta_k=\theta_k^{(0)}$ against $\theta_k=\theta_k^{(1)}$, where the boundary values are set such that the error constraints are satisfied at $\theta_k^{(0)}, \theta_k^{(1)}$. For some important cases, such as an exponential family of distributions, this sequential test has the property that type $I$ and type $II$ errors are less than $\alpha_k$, $\beta_k$ for all $\theta_k<\theta_k^{(0)}$ and $\theta_k>\theta_k^{(1)}$, respectively. However, while the SPRT minimizes the expected sample size at $\theta_k=\theta_k^{(0)}, \theta_k^{(1)}$, it is highly sub-optimal for other values of $\theta$, as demonstrated in Section \ref{sec:simulation}. Therefore, other techniques should be considered under the composite hypothesis case.

Let $\boldsymbol\theta_k$ be a vector of unknown parameters of component $k$. The observations $\left\{y_k(i)\right\}_{i\geq1}$ are drawn from a common distribution $f\left(y|\boldsymbol\theta_k\right)$, $\boldsymbol\theta_k\in\Theta_k$, where $\Theta_k$ is the parameter space of component $k$. If component $k$ is healthy, then $\boldsymbol\theta_k\in\Theta_k^{(0)}$; if component $k$ is abnormal, then $\boldsymbol\theta_k\in(\Theta\backslash\Theta_k^{(0)})$. Let $\Theta_k^{(0)}$, $\Theta_k^{(1)}$ be disjoint subsets of $\Theta_k$, where $I_k=\Theta\backslash(\Theta_k^{(0)}\cup\Theta_k^{(1)})\neq\emptyset$ is an indifference region\footnote{The assumption of an indifference region is widely used in the theory of sequential testing of composite hypotheses to derive asymptotically optimal performance. Nevertheless, in some cases this assumption can be removed. For more details, the reader is referred to \cite{Lai_1988_Nearly}.}.
When $\boldsymbol\theta_k\in I_k$, the detector is indifferent regarding the state of component $k$. Hence, there are no constraints on the error probabilities for all $\boldsymbol\theta_k\in I_k$. The hypothesis test regarding component $k$ is to test
\begin{center}
$\boldsymbol\theta_k\in\Theta_k^{(0)}$ \; against \; $\boldsymbol\theta_k\in\Theta_k^{(1)}$.
\end{center}
Narrowing $I_k$ has the price of increasing the sample size. \\
Let
\beq
\bea{l}
\hat{\boldsymbol\theta}_k(n)
=\displaystyle\arg\max_{\theta_k\in\Theta_k}
                        {f\left(\mathbf{y}_k(n)|\boldsymbol\theta_k\right)} , \vspace{0.1cm}\\
\hat{\boldsymbol\theta}_k^{(i)}(n)
=\displaystyle\arg\max_{\theta_k\in\Theta_k^{(i)}}
                        {f\left(\mathbf{y}_k(n)|\boldsymbol\theta_k\right)} ,
\ena
\eeq
be the Maximum-Likelihood Estimates (MLEs) of the parameters over the parameter spaces $\Theta_k$, $\Theta_k^{(i)}$ at stage $n$, respectively.

In contrast to the SPRT (for the simple hypothesis case), the theory of sequential tests of composite hypotheses does not provide optimal performance in terms of minimizing the expected sample size under given error constraints. Nevertheless, asymptotically optimal performance can be obtained as the error probability approaches zero.

First, we provide an overview of existing sequential tests for composite hypotheses which are relevant to our problem. Next, we apply these techniques to solve (\ref{eq:opt1}).

\subsection{Existing Sequential Tests for Composite Hypothesis Testing}
\label{ssec:existing}

The key idea is to use the estimated parameters to perform a one-sided sequential test to reject $H_0$ and a one-sided sequential test to reject $H_1$.
Note that these techniques were introduced for a single process. However, in this paper we apply sequential tests for $K$ components. Thus, we use the subscript $k$ to denote the component index.

\subsubsection{Sequential Generalized Likelihood Ratio Test (SGLRT)}

We refer to sequential tests that use the Generalized Likelihood Ratio (GLR) statistics as the SGLRT. \\
For $i=0, 1$, let
\beq
\label{eq:GLR}
L_k^{(i), GLR}(n)=\displaystyle\log\frac{\prod_{r=1}^{n}{f(y_k(r)|\hat{\boldsymbol\theta}_k(n))}}
                        {\prod_{r=1}^{n}{f(y_k(r)|\hat{\boldsymbol\theta}_k^{(i)}(n))}}
\eeq
be the GLR statistics used to reject hypothesis $H_i$ at stage $n$.  \\
Let
\beq\label{eq:GLR_stopping}
\bea{l}
N_k^{(i)}=\displaystyle\inf\left\{ \; n \; :  L_k^{(i), GLR}(n) \geq B_k^{(i)}
\right\} \;,
\ena
\eeq
be the stopping rule used to reject hypothesis $H_i$. $B_k^{(i)}$ is the boundary value. \\
For each component $k$, the decision maker stops the sampling when $N_k=\min\left\{N_k^{(0)}, N_k^{(1)}\right\}$. If $N_k=N_k^{(0)}$, component $k$ is declared as abnormal (i.e., $H_0$ is rejected). If $N_k=N_k^{(1)}$, component $k$ is declared as normal (i.e., $H_0$ is accepted).

The SGLRT was first studied by Schwartz \cite{Schwarz_1962_Asymptotic} for a one-parameter exponential family, who assigned a cost of $c$ for each observation and a loss function for wrong decisions. It was shown that setting $B_k^{(i)}=\log(c^{-1})$ asymptotically minimizes the Bayes risk as $c$ approaches zero. A refinement was studied by Lai \cite{Lai_1988_Nearly, Lai_1994_Nearly}, who set a time-varying boundary value $B_k^{(i)}\sim\log((nc)^{-1})$. Lai showed that for a multivariate exponential family this scheme asymptotically minimizes both the Bayes risk and the expected sample size subject to error constraints as $c$ approaches zero \cite{Lai_1994_Nearly}.

\subsubsection{Sequential Adaptive Likelihood Ratio Test (SALRT)}

We refer to sequential tests that use the Adaptive Likelihood Ratio (ALR) statistics as the SALRT. \\
For $i=0, 1$, let
\beq
\label{eq:ALR}
L_k^{(i), ALR}(n)=\displaystyle\log\frac{\prod_{r=1}^{n}{f(y_k(r)|\hat{\boldsymbol\theta}_k(r-1))}}
                         {\prod_{r=1}^{n}{f(y_k(r)|\hat{\boldsymbol\theta}_k^{(i)}(n))}}
\eeq
be the ALR statistics used to reject hypothesis $H_i$ at stage $n$. \\
Let
\beq\label{eq:ALR_stopping}
\bea{l}
N_k^{(i)}=\displaystyle\inf\left\{ \; n \; :  L_k^{(i), ALR}(n) \geq B_k^{(i)}
\right\} \;,
\ena
\eeq
be the stopping rule used to reject hypothesis $H_i$, where $B_k^{(i)}$ is the boundary value.\\
For each component $k$, the decision maker stops the sampling when $N_k=\min\left\{N_k^{(0)}, N_k^{(1)}\right\}$. If $N_k=N_k^{(0)}$, component $k$ is declared as abnormal. If $N_k=N_k^{(1)}$, component $k$ is declared as normal.

The SALRT was first introduced by Robbins and Siegmund \cite{Robbins_1974_Expected} to design power-one sequential tests. Pavlov used it to design asymptotically (as the error probability approaches zero) optimal (in terms of minimizing the expected sample size subject to error constraints) tests for composite hypothesis testing of the multivariate exponential family \cite{Pavlov_1990_Sequential}. Tartakovsky established asymptotically optimal performance for a more general multivariate family of distributions \cite{Tartakovsky_2002_Efficient}.

The advantage of using the SALRT is that setting $B_k^{(0)}=\log\frac{1}{\alpha_k}$, $B_k^{(1)}=\log\frac{1}{\beta_k}$ satisfies the error probability constraints in (\ref{eq:sub_k_opt}). However, such a simple setting cannot be applied to the SGLRT. Thus, implementing the SALRT is much simpler than implementing the SGLRT. The disadvantage of using the SALRT is that poor early estimates (for small number of observations) can never be revised even though one has a large number of observations.

\subsection{Asymptotically Optimal Index Policies}
\label{ssec:modifying}

It is intuitive that the selection rules in the composite hypothesis case remain the same as in the simple hypothesis case. The resulting strategies are thus referred to as $\pi c N$-SGLRT/SALRT and $\pi c N_0$-SGLRT/SALRT algorithms. In the following theorems, we show that the $\pi c N$-SGLRT/SALRT and $\pi c N_0$-SGLRT/SALRT algorithms are asymptotically optimal in terms of minimizing the objective function subject to the error constraints (\ref{eq:opt1}) as the error probabilities approach zero\footnote{As shown in the proof of Theorems \ref{th:asymptotic3}, \ref{th:asymptotic4}, the index policies are still optimal in terms of testing order. The asymptotic optimality is due to the performance of the sequential test under the composite hypothesis case.}. When deriving asymptotics we assume that $P_k^{FA}\rightarrow 0, P_k^{MD}\rightarrow 0$ for all $k$ such that the asymptotic optimality property in terms of minimizing the expected sample size subject to the error constraints holds for each single process for both SGLRT and SALRT, as discussed in Section \ref{ssec:existing}.

\textsl{\theorem\label{th:asymptotic3}{
Consider the independent model under the composite hypothesis case. Let $(\boldsymbol\tau^{OPT}, \boldsymbol\delta^{OPT}, \boldsymbol\phi^{OPT})$ be the optimal solution to (\ref{eq:opt1}). Let $(\boldsymbol\tau^{*}, \boldsymbol\delta^{*}, \boldsymbol\phi^{*})$ be the solution achieved by the $\pi c N$-SGLRT/SALRT algorithm. Then, as $P_k^{FA}\rightarrow 0, P_k^{MD}\rightarrow 0$ for all $k$, we obtain:
\beq
\bea{l}
\E\left\{\displaystyle\sum_{k\in \mathcal{H}_1}{c_k\tau_k}|(\boldsymbol\tau^{*}, \boldsymbol\delta^{*}, \boldsymbol\phi^{*})\right\} \vspace{0.1cm} \\ \hspace{2cm}
\sim
\E\left\{\displaystyle\sum_{k\in \mathcal{H}_1}{c_k\tau_k}|(\boldsymbol\tau^{OPT}, \boldsymbol\delta^{OPT}, \boldsymbol\phi^{OPT})\right\}
\ena
\eeq
}} \\
\begin{proof}
See Appendix \ref{app:comp_ind}.
\end{proof}
\textsl{\theorem\label{th:asymptotic4}{
Consider the exclusive model under the composite hypothesis case. Let $(\boldsymbol\tau^{OPT}, \boldsymbol\delta^{OPT}, \boldsymbol\phi^{OPT})$ be the optimal solution to (\ref{eq:opt1}). Let $(\boldsymbol\tau^{*}, \boldsymbol\delta^{*}, \boldsymbol\phi^{*})$ be the solution achieved by the $\pi c N_0$-SGLRT/SALRT algorithm. Then, as $P_k^{FA}\rightarrow 0, P_k^{MD}\rightarrow 0$ for all $k$, we obtain:
\beq
\bea{l}
\E\left\{\displaystyle\sum_{k\in \mathcal{H}_1}{c_k\tau_k}|(\boldsymbol\tau^{*}, \boldsymbol\delta^{*}, \boldsymbol\phi^{*})\right\} \vspace{0.1cm} \\ \hspace{2cm}
\sim
\E\left\{\displaystyle\sum_{k\in \mathcal{H}_1}{c_k\tau_k}|(\boldsymbol\tau^{OPT}, \boldsymbol\delta^{OPT}, \boldsymbol\phi^{OPT})\right\}
\ena
\eeq
}} \\
\begin{proof}
See Appendix \ref{app:comp_exc}.
\end{proof}

\subsection{Computing the Indices}
\label{ssec:computing2}

Arranging the components in decreasing order of $\pi_kc_k/\mathbf{E}(N_k)$ or $\pi_kc_k/\mathbf{E}(N_k|H_0)$ requires one to compute the expected sample size $\mathbf{E}(N_k|H_i)$ for all $k=1, 2, ..., K$. In general, it is difficult to obtain a closed-form expression for the exact value of $\mathbf{E}(N_k|H_i)$. However, we can use the asymptotic property of the tests to obtain a closed-form approximation of $\mathbf{E}(N_k|H_i)$, which approaches the exact expected sample size as the error probability approaches zero.

For every $i=0 ,1$, let
\beq
D_k(\boldsymbol\theta_k||\boldsymbol\lambda)
=\mathbf{E}_{\boldsymbol\theta_k}
   \left(\log\frac{f(y_k(1)|\boldsymbol\theta_k)}{f(y_k(1)|\boldsymbol\lambda)}\right)
\eeq
be the KL divergence between the real value of $\boldsymbol\theta_k$ and $\boldsymbol\lambda$, where the expectation is taken with respect to $f(y|\boldsymbol\theta_k)$,\\
and let
\beq
D_k^*(\boldsymbol\theta_k||\Theta_k^{(i)})
=\displaystyle\inf_{\boldsymbol\lambda\in\Theta_k^{(i)}} D_k(\boldsymbol\theta_k||\boldsymbol\lambda)\;.
\eeq
Let $I_k^{(0)}, I_k^{(1)}$ be disjoint subsets of $I_k$ and $I_k=I_k^{(0)}\cup I_k^{(1)}$, such that for all $\boldsymbol\theta_k\in I_k^{(i)}$ we have $\frac{B_k^{(j)}}{D_k^*(\boldsymbol\theta_k||\Theta_k^{(j)})}\leq \frac{B_k^{(i)}}{D_k^*(\boldsymbol\theta_k||\Theta_k^{(i)})}$ for $i, j = 0, 1$. Let $P^{(i)}(\boldsymbol\theta_k)$ be a prior distribution on $\boldsymbol\theta_k$ over $\Theta_k^{(i)}\cup I_k^{(i)}$ (corresponding to $H_i$). Then, as $P_k^{FA}\rightarrow 0, P_k^{MD}\rightarrow 0$, the conditional expected sample size is given by \cite{Pavlov_1990_Sequential, Lai_1994_Nearly}:
\beq
\label{eq:sample_size_approx_uncertainty}
\bea{l}
\mathbf{E}(N_k|H_0)\sim\displaystyle\int_{\boldsymbol\theta_k\in\Theta_k^{(0)}\cup I_k^{(0)}}{ \frac{B_k^{(1)}}
            {D_k^*(\boldsymbol\theta_k||\Theta_k^{(1)})}dP^{(0)}(\boldsymbol\theta_k)} \;,
\vspace{0.1cm} \\
\mathbf{E}(N_k|H_1)\sim %\vspace{0.1cm} \\
\displaystyle\int_{\boldsymbol\theta_k\in\Theta_k^{(1)}\cup I_k^{(1)}}{ \frac{B_k^{(0)}}
            {D_k^*(\boldsymbol\theta_k||\Theta_k^{(0)})}dP^{(1)}(\boldsymbol\theta_k)} \;.
\ena
\eeq
The expected sample size required to make a decision regarding the state of component $k$ is given by:
\beq
\bea{l}
\mathbf{E}(N_k) \vspace{0.3cm}
= \pi_k\mathbf{E}(N_k|H_1)+(1-\pi_k)\mathbf{E}(N_k|H_0) \;,
\ena
\eeq
which can be well approximated for small error probability using (\ref{eq:sample_size_approx_uncertainty}).

\section{Numerical Examples}
\label{sec:simulation}

In this section we present numerical examples to illustrate the performance of the algorithms. Consider a cyber network consisting of $K$ components (which can be routers, paths, etc.), as discussed in Section \ref{sec:intro_app}. Assume that an intruder tries to launch a DoS or Reduction of Quality (RoQ) attacks by sending a large number of packets to a component. RoQ attacks inflict damage on the component, while keeping a low profile to avoid detection. RoQ attacks do not cause denial of service.

To detect such attacks, the IDS performs a traffic-based anomaly detection. It monitors the traffic at each component to decide whether a component is compromised. Roughly speaking, if the actual arrival rate is significantly higher than the arrival rate under the normal state, then the IDS should declare that the component is in an abnormal state. A similar traffic-based detection technique was proposed in \cite{Onat_2005_Intrusion} for a different model, considering a single process without switching to other components. For each component $k$, we assume that packets arrive according to a Poisson process with rate $\theta^{(k)}$.
When component $k$ is tested, the IDS collects an observation $y_k(n)\in\mathbb{N}_0$ every time unit, which represents the number of packets that arrived in the interval $(n-1, n)$. Assume that the IDS considers component $k$ as normal if $\theta_k\leq\theta_k^{(0)}$, and tests $\theta_k\leq\theta_k^{(0)}$ against $\theta_k\geq\theta_k^{(1)}$ (i.e., $I_k=\{\theta_k|\theta_k^{(0)}<\theta_k<\theta_k^{(1)}\}$ is the indifference region). We set $c_k=\theta_k^{(0)}$. As discussed in Section \ref{sec:intro_app}, under this setting the optimization problem minimizes the maximal damage to the network in terms of packet-loss.

\subsection{Simple Hypothesis Case}

We consider the case where the observations follow Poisson distributions $y_k(n)\sim\mathrm{Poi}(\theta_k^{(0)})$ or $y_k(n)\sim\mathrm{Poi}(\theta_k^{(1)})$ depending on wether component $k$ is healthy or abnormal, respectively, where $\theta_k^{(0)}, \theta_k^{(1)}$ are known to the IDS. To implement the $\pi c N$-SPRT and $\pi c N_0$-SPRT algorithms (which are optimal in this scenario for the independent and exclusive models, respectively), we need to compute the LR between the hypotheses, defined in (\ref{eq:LR}), and the expected sample sizes under the hypotheses, which can be well approximated by (\ref{eq:sample_size_approx_H}). Let $\Lambda_k(n)=\log L_k(n)$ be the Log-Likelihood Ratio (LLR) between the two hypotheses of component $k$ at stage $n$, where $L_k(n)$ is defined in (\ref{eq:LR}). After algebraic manipulations, it can be verified that the LLR is given by:
\beq
\displaystyle\Lambda_k(n)=-n\left(\theta_k^{(1)}-\theta_k^{(0)}\right)
                            +\log\left(\theta_k^{(1)}/\theta_k^{(0)}\right)\sum_{i=1}^{n}{y_k(i)} \;.
\eeq
It can be verified that the KL divergence between the hypotheses $H_i$ and $H_j$, defined in (\ref{eq:KL}), is given by:
\beq
\label{eq:KL_Poisson}
D_k(i||j)=\theta_k^{(j)}-\theta_k^{(i)}+\theta_k^{(i)}\log\left(\theta_k^{(i)}/\theta_k^{(j)}\right) \;.
\eeq
Substituting (\ref{eq:KL_Poisson}) in (\ref{eq:sample_size_approx_H}) yields the required approximation to the expected sample size. We note that the optimal indices order was preserved using the approximation in (\ref{eq:sample_size_approx_H}) under all numerical examples in this section.

Next, we provide numerical examples to illustrate the performance of the algorithms. We compared three schemes: a Random selection SPRT (R-SPRT), where a series of SPRTs are performed until all the components are tested in a random order (which is optimal for the problem of minimizing the detection delay over independent processes \cite{Caromi_2013_Fast}), and the proposed $\pi c N$-SPRT and $\pi c N_0$-SPRT algorithms, which are optimal under the independent and exclusive models, respectively.

\begin{figure}[t] %[htbp]
\begin{center}
    \subfigure[An independent model scenario.]{\scalebox{0.4}
    {
      \label{fig:fig1a}{\epsfig{file=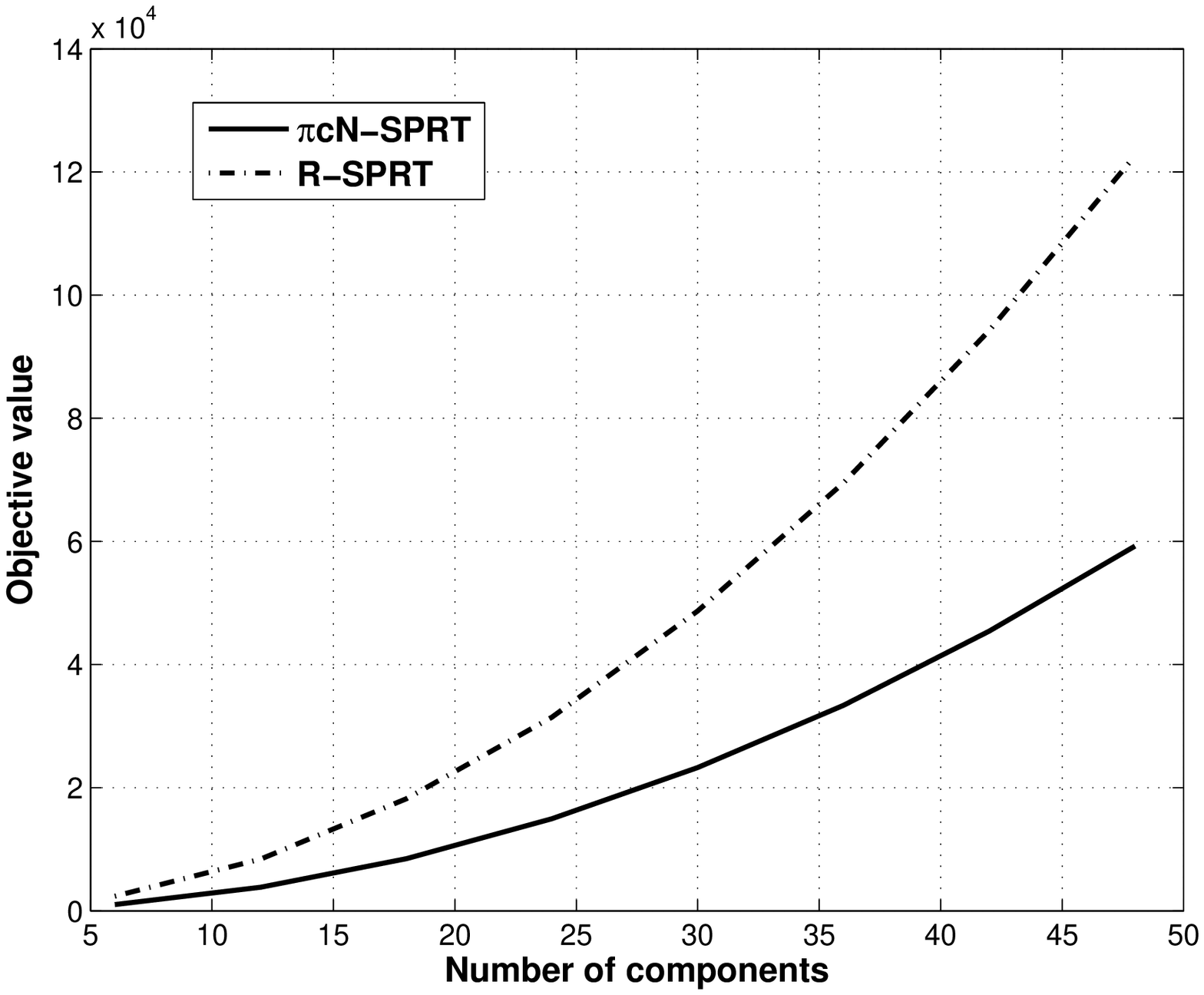}}
    }}
    \subfigure[An exclusive model scenario.]{\scalebox{0.4}
    {
      \label{fig:fig1b}{\epsfig{file=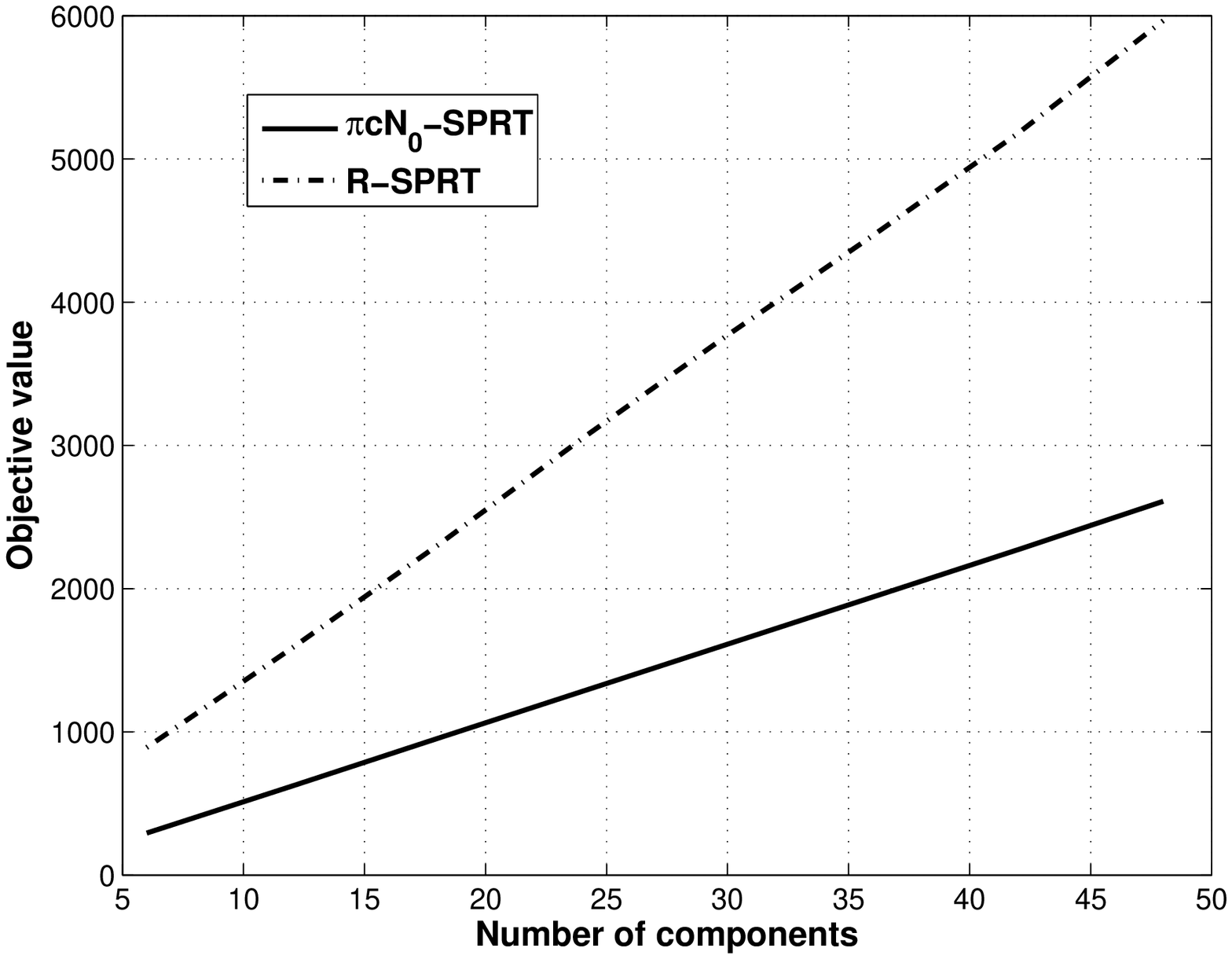}}
    }}
   \caption{Objective value as a function of the number of components under the independent and exclusive models.}
  \label{fig:fig1}
\end{center}
  \end{figure}

Let $\Delta_K=(100-10)/(K-1)$. We set $c_k=\theta_k^{(0)}=10+(k-1)\Delta_K$ (i.e., the costs are equally spaced in the interval $[10 , 100]$) and $\theta_k^{(1)}=1.5\cdot\theta_k^{(0)}$. The error constraints were set to $P_k^{FA}=10^{-2}, P_k^{MD}=10^{-6}$ for all $k$. For the independent and exclusive models, we set $\pi_k=0.8$ and $\pi_k=1/K$ for all $k$, respectively. The performance of the $\pi c N$-SPRT and $\pi c N_0$-SPRT algorithms are presented in Fig. \ref{fig:fig1a} and \ref{fig:fig1b} under the independent and exclusive models, respectively, and compared to the R-SPRT. It can be seen that the proposed algorithms save roughly $50\%$ of the objective value as compared to the R-SPRT under both the independent and exclusive model scenarios.

\begin{figure}[htbp]
\centering \epsfig{file=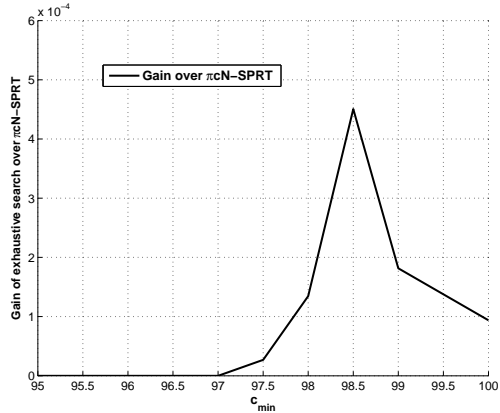,
width=0.4\textwidth}
\caption{Performance gain of an exhaustive search over the $\pi c N$-SPRT algorithm as a function of $c_{min}$ under the independent model.}
\label{fig:fig_multiple}
\end{figure}

Next, we simulate the independent model when $2$ components are observed at a time and the total number of components is $K=6$. Note that in this case the $\pi c N$-SPRT algorithm may not be optimal. We use an exhaustive search as a bench mark to demonstrate the performance of the $\pi c N$-SPRT algorithm in this scenario. The exhaustive search is done by performing a sequence of $K$ SPRTs among all the possible testing orders. Then, the minimal objective value is chosen as a bench mark. We set the maximal cost to $c_{max}=100$ and the costs are equally spaced in the interval $[c_{min} , 100]$. The error constraints were set to $P_k^{FA}=P_k^{MD}=10^{-2}$ for all $k$. The performance gain of the exhaustive search scheme over the $\pi c N$-SPRT algorithm as a function of $c_{min}$ are presented in Fig. \ref{fig:fig_multiple}. It can be seen that the $\pi c N$-SPRT algorithm almost achieves the performance of the exhaustive search scheme in this scenario for all $c_{min}$. For small $c_{min}$ both algorithms perform the same, since the difference between the indices increases. The exhaustive search outperforms the $\pi c N$-SPRT algorithm for $c_{min}>97$, but the gain remains very small.
\subsection{Composite Hypothesis Case}

We consider the case of composite hypotheses, where there is uncertainty in the distribution parameters, as discussed in Section \ref{sec:uncertainty}. To implement the asymptotically optimal the $\pi c N$-SGLRT/SALRT and $\pi c N_0$-SGLRT/SALRT algorithms, we need to compute the GLR or ALR statistics, defined in (\ref{eq:GLR}), (\ref{eq:ALR}) and the expected sample sizes under the hypotheses, which can be well approximated by (\ref{eq:sample_size_approx_uncertainty}). The MLEs of the parameters over the parameter spaces $\Theta_k$, $\Theta_k^{(i)}$ are given by the sample mean and the boundary of the alternative parameter space, respectively. As a result, substituting: $\hat{\theta}_k(n)
=\frac{1}{n}\sum_{i=1}^{n}{y_k(i)}  \;,
\hat{\theta}_k^{(i)}(n)
=\theta_k^{(i)}$ ,
in (\ref{eq:GLR}), (\ref{eq:ALR}) yields the GLR and ALR statistics, respectively. The KL divergence between the real value of $\theta_k$ and the parameter space $\Theta_k^{(i)}$ is given by:
\beq
\label{eq:KL_Poisson_uncertainty}
D_k^*(\theta_k||\Theta_k^{(i)})
=\theta_k^{(i)}-\theta_k+\theta_k\log\left(\theta_k/\theta_k^{(i)}\right) \;.
\eeq
Substituting (\ref{eq:KL_Poisson_uncertainty}) in (\ref{eq:sample_size_approx_uncertainty}) yields the approximate expected sample size.

\begin{figure}[htbp]
\centering \epsfig{file=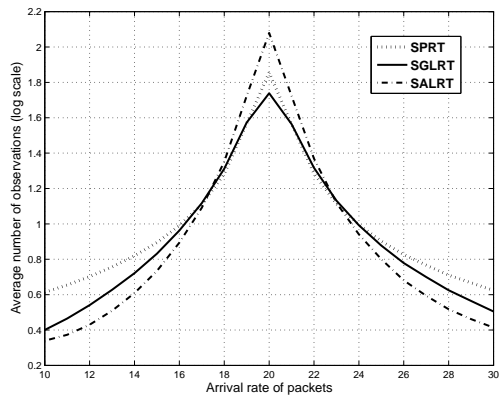,
width=0.4\textwidth}
\caption{Average number of observations as a function of the arrival rate of packets (denoted by $\theta$).}
\label{fig:fig2}
\end{figure}

Next, we provide numerical examples to illustrate the performance of the algorithms under uncertainty.
We simulated a network with homogenous components (i.e., any selection rule is optimal). We compared three schemes: R-SPRT, and the $\pi c N$-SGLRT/SALRT or $\pi c N_0$-SGLRT/SALRT algorithms (which achieve the same performance in this case) using the SALRT and the SGLRT, discussed in section \ref{ssec:existing}. We set $\theta_k^{(0)}=19$, $\theta_k^{(1)}=21$. Under uncertainty, the IDS considers component $k$ as normal if $\theta_k\leq\theta_k^{(0)}$, and tests $\theta_k\leq\theta_k^{(0)}$ against $\theta_k\geq\theta_k^{(1)}$ (i.e., $I_k=\{\theta_k|19<\theta_k<21\}$ is the indifference region). To implement the SGLRT, we set the cost per observation $c=10^{-3}$. According to the assigned cost, we obtained the following error probability constraints for all $k$: $P_k^{FA}\leq 0.026$ for all $\theta^{(k)}\leq 19$ and $P_k^{MD}\leq 0.03$ for all $\theta^{(k)}\geq 21$. We do not restrict the detector's performance for $19<\theta^{(k)}<21$ (Note that narrowing the indifference region has the price of increasing the required sample size). In Fig. \ref{fig:fig2} we show the average number of observations (in a log scale) required for the anomaly detection as a function of $\theta^{(k)}$. As expected, for $\theta_k=19$ and $\theta_k=21$ the R-SPRT requires lower sample size as compared to the proposed schemes. On the other hand, it can be seen that for most values of $\theta$ the SGLRT and the SALRT require lower sample size as compared to the R-SPRT. The SALRT performs the worst for $18<\theta_k<22$, and performs the best for $\theta_k\nin (18,22)$, roughly. The SGLRT obtains the best average performance. It can be seen that for large values of $\theta_k$ the anomaly is detected very quickly, since the distance between the hypotheses increases. This result confirms that DoS attacks are much easier to detect than RoQ attacks.

\section{Conclusion}
\label{sec:conclusion}
The problem of anomaly localization in a resource-constrained cyber system was studied. Due to resource constraints, only one component can be probed at a time. The observations are realizations drawn from two different distributions depending on whether the component is normal or anomalous. An abnormal component incurs a cost per unit time until it is tested and identified. The problem was formulated as a constrained optimization problem. The objective is to minimize the total expected cost subject to error probability constraints. We considered two different anomaly models: the independent model in which each component can be abnormal independent of other components, and the exclusive model in which there is one and only one abnormal component. For the simple hypothesis case, we derived optimal algorithms for both independent and exclusive models. For the composite hypothesis case, we derived asymptotically (as the error probability approaches zero) optimal algorithms for both independent and exclusive models. These optimal algorithms have low-complexity.

The algorithms developed in this paper can be applied to other models of anomaly detection as well. We can modify the proposed algorithms to any detection scheme that performs a series of tests according to the $\pi c N$-rule or $\pi c N_0$-rule. The required modification is to replace the SPRT/SALRT/SGLRT by any given test. Such modified algorithms minimize the objective function among all the algorithms that perform the given test.

Deriving optimal policies for the anomaly localization problem considered in this paper requires the assumption that switching to a different component is allowed only when the state of the currently probed component is declared. A future research direction is to examine the anomaly localization problem under the case where switching to a different component and declarations of the states of individual components are allowed at all times.

\section{Appendix}
\label{app}

In this appendix we provide the proofs for Theorems $1-3$. For convenience, we use the superscripts $A1, A2$ when referring to the $\pi c N$-SPRT and $\pi c N_0$-SPRT algorithms, respectively. We use the superscripts $A3, A4$ when referring to the $\pi c N$-SGLRT/SALRT and $\pi c N_0$-SGLRT/SALRT algorithms, respectively.

Throughout the proofs, we use the specific formula for the updated posterior probability of component $k$ being abnormal.

Let $\mathbf{1}_k(n)$ be the probing indicator function, where $\mathbf{1}_k(n)=1$ if component $k$ is probed at time $n$ and $\mathbf{1}_k(n)=0$ otherwise. Let $t_m$ be the time when the decision maker starts the $m^{th}$ test. For example, assume that $K=3$ and the decision maker tests the components according to the following order: $3, 1, 2$. Then, $t_1=1$ (when the test starts), $t_2=\tau_3+1$, $t_3=\tau_1+1$.

Under the independent model, the posterior probability of component $k$ being abnormal can be updated at time $t_{m+1}$ as follows \cite{Castanon_1995_Optimal}:
\beq
\label{eq:beleif_ind}
\bea{l}
\pi_k(t_{m+1})=\displaystyle\left(1-\mathbf{1}_k(t_m)\right)\pi_k(t_m) \vspace{0.1cm}\\ \hspace{0.5cm}
+\displaystyle\frac{\mathbf{1}_k(t_m)\pi_k(t_m)f_k^{(1)}(\mathbf{y}_k(N_k))}
                {\pi_k(t_m)f_k^{(1)}(\mathbf{y}_k(N_k))+\left(1-\pi_k(t_m)\right)f_k^{(0)}(\mathbf{y}_k(N_k))}\;,
\ena
\eeq
where $\pi_k(t_1)=\pi_k$ denotes the \emph{a priori} probability of component $k$ being abnormal. The term $\mathbf{y}_k(N_k)=\left\{y_k(i)\right\}_{i=t_m}^{t_m+N_k-1}$ denotes the $N_k$-size vector of observations, taken from component $k$. Under the exclusive model, $\pi_k(t_{m+1})$ is given in (\ref{eq:beleif_exc}) at the top of the next page. Note that in contrast to the independent model, under the exclusive model the beliefs of all the components are changed at each time due to the dependency across components. The posterior probabilities depend on the selection rule and the collected measurements.
\begin{figure*}[!t]
\normalsize
\begin{equation}
\label{eq:beleif_exc}
\bea{l}
\pi_k(t_{m+1})=
\displaystyle\frac{\mathbf{1}_k(t_m)\pi_k(t_m)f_k^{(1)}(\mathbf{y}_k(N_k))}
                {\pi_k(t_m)f_k^{(1)}(\mathbf{y}_k(N_k))+\left(1-\pi_k(t_m)\right)f_k^{(0)}(\mathbf{y}_k(N_k))}
                \vspace{0.1cm}\\ \hspace{1.8cm}
+\displaystyle\frac{\left(1-\mathbf{1}_k(t_m)\right)\pi_k(t_m)f_{\phi(t_m)}^{(0)}(\mathbf{y}_{\phi(t_m)}(N_{\phi(t_m)}))}
                {\pi_{\phi(t_m)}(t_m)f_{\phi(t_m)}^{(1)}(\mathbf{y}_{\phi(t_m)}(N_{\phi(t_m)}))
                +\left(1-\pi_{\phi(t_m)}(t_m)\right)f_{\phi(t_m)}^{(0)}(\mathbf{y}_{\phi(t_m)}(N_{\phi(t_m)}))}
 \;.
\ena
\end{equation}
\hrulefill
\vspace*{4pt}
\end{figure*}

\subsection{Proof of Theorem \ref{th:optimality_alg1_2} Under The Exclusive Model}
\label{app:exclusive}
Let $\mathbf{E}'(N_k|H_i,$t$)$ be the expected sample size achieved by a stopping rule and a decision rule $(\tau'_k(t), \delta'_k(t))$, depending on the time that component $k$ is tested (i.e., $(\tau'_k(t), \delta'_k(t))$ depend on the selection rule), such that error constraints are satisfied. Let $\mathbf{E}^{A2}(N_k|H_i)$ be the expected sample size achieved by the SPRT's stopping rule and decision rule $(\tau^{A2}_k, \delta^{A2}_k)$, independent of the time that component $k$ is tested (i.e., $(\tau^{A2}_k, \delta^{A2}_k)$ are independent of the selection rule), such that error constraints are satisfied. Clearly, $\mathbf{E}^{A2}(N_k|H_i)\leq \mathbf{E}'(N_k|H_i,t)$ for all $k ,t$, for $i=0, 1$. \vspace{0.1cm}\\
\emph{\textbf{Step $1$:} Proving the theorem for $K=2$:} \vspace{0.1cm}\\
Assume that
\beq
\frac{\pi_1(t_1)c_1}{\mathbf{E}^{A2}(N_1|H_0)}\geq\frac{\pi_2(t_1)c_2}{\mathbf{E}^{A2}(N_2|H_0)} \;.
\eeq
Consider selection rules $\boldsymbol\phi^{(1)}$, $\boldsymbol\phi^{(2)}$ that select component $1$ first followed by component $2$ and component $2$ first followed by component $1$, respectively. The expected cost achieved by $(\boldsymbol\tau'(t), \boldsymbol\delta'(t), \boldsymbol\phi^{(2)})$ is given by:
\beq
\bea{l}
\E\left\{\displaystyle\sum_{k=1}^{K}{c_k \tau_k \mathbf{1}_{\left\{k\in \mathcal{H}_1\right\}}}\;|\;(\boldsymbol\tau'(t), \boldsymbol\delta'(t), \boldsymbol\phi^{(2)})\right\} \vspace{0.1cm}\\ \hspace{0.5cm}
=\left(\mathbf{E}'(N_2|H_1, t_1)\right)\pi_2(t_1)c_2 \vspace{0.1cm}\\ \hspace{1cm}
+ \left(\mathbf{E}'(N_2|H_0, t_1) + \mathbf{E}'(N_1|H_1, t_2)\right)\pi_1(t_1) c_1.
\ena
\eeq
The expected cost achieved by $(\boldsymbol\tau'(t), \boldsymbol\delta'(t), \boldsymbol\phi^{(1)})$ is given by:
\beq
\bea{l}
\E\left\{\displaystyle\sum_{k=1}^{K}{c_k \tau_k \mathbf{1}_{\left\{k\in \mathcal{H}_1\right\}}}\;|\;(\boldsymbol\tau'(t), \boldsymbol\delta'(t), \boldsymbol\phi^{(1)})\right\} \vspace{0.1cm}\\ \hspace{0.5cm}
=\left(\mathbf{E}'(N_1|H_1, t_1)\right)\pi_1(t_1)c_1 \vspace{0.1cm}\\ \hspace{1cm}
+ \left(\mathbf{E}'(N_1 | H_0, t_1) + \mathbf{E}'(N_2|H_1, t_2)\right)\pi_2(t_1) c_2.
\ena
\eeq
Note that the expected cost achieved by both selection rules can be further reduced by minimizing the expected sample sizes (such that error constraints are satisfied) independent of the selection rules, which is achieved by $(\tau^{A2}_k, \delta^{A2}_k)$. Therefore, an optimal solution must be $(\boldsymbol\tau^{A2}, \boldsymbol\delta^{A2}, \boldsymbol\phi^{(1)})$ or $(\boldsymbol\tau^{A2}, \boldsymbol\delta^{A2}, \boldsymbol\phi^{(2)})$. Next, we use the interchange argument to prove the theorem for $K=2$. The expected cost achieved by $(\boldsymbol\tau^{A2}, \boldsymbol\delta^{A2}, \boldsymbol\phi^{(2)})$ is given by:
\beq
\bea{l}
\E\left\{\displaystyle\sum_{k=1}^{K}{c_k \tau_k \mathbf{1}_{\left\{k\in \mathcal{H}_1\right\}}}\;|\;(\boldsymbol\tau^{A2}, \boldsymbol\delta^{A2}, \boldsymbol\phi^{(2)})\right\} \vspace{0.1cm}\\ \hspace{0.5cm}
=\left(\mathbf{E}^{A2}(N_2|H_1)\right)\pi_2(t_1)c_2 \vspace{0.1cm}\\ \hspace{1cm}
+ \left(\mathbf{E}^{A2}(N_2|H_0) + \mathbf{E}^{A2}(N_1|H_1)\right)\pi_1(t_1) c_1.
\ena
\eeq
The expected cost achieved by $(\boldsymbol\tau^{A2}, \boldsymbol\delta^{A2}, \boldsymbol\phi^{(1)})$ is given by:
\beq
\bea{l}
\E\left\{\displaystyle\sum_{k=1}^{K}{c_k \tau_k \mathbf{1}_{\left\{k\in \mathcal{H}_1\right\}}}\;|\;(\boldsymbol\tau^{A2}, \boldsymbol\delta^{A2}, \boldsymbol\phi^{(1)})\right\} \vspace{0.1cm}\\ \hspace{0.5cm}
=\left(\mathbf{E}^{A2}(N_1|H_1)\right)\pi_1(t_1)c_1 \vspace{0.1cm}\\ \hspace{1cm}
+ \left(\mathbf{E}^{A2}(N_1|H_0) + \mathbf{E}^{A2}(N_2|H_1)\right)\pi_2(t_1) c_2.
\ena
\eeq
the expected cost achieved by $\boldsymbol\phi^{(1)}$ is lower than that achieved by $\boldsymbol\phi^{(2)}$ since $\displaystyle\frac{\pi_1(t_1)c_1}{\mathbf{E}^{A2}(N_1|H_0)}\geq\frac{\pi_2(t_1)c_2}{\mathbf{E}^{A2}(N_2|H_0)}$, which completes the proof for $K=2$.\vspace{0.1cm}\\
\emph{\textbf{Step $2$:} Proving the theorem by induction on the number of components $K$:} \vspace{0.1cm}\\
Assume that the theorem is true for $K-1$ components (where one and only one component is abnormal).
Assume that
\beq
\frac{\pi_1(t_1)c_1}{\mathbf{E}^{A2}(N_1|H_0)}\geq\frac{\pi_2(t_1)c_2}{\mathbf{E}^{A2}(N_2|H_0)}\geq ... \geq\frac{\pi_K(t_1)c_K}{\mathbf{E}^{A2}(N_K|H_0)} \;.
\eeq
Consider the case of $K$ components and denote $\boldsymbol\phi^{(j)}$ as an optimal selection rule  that selects component $j$ first. \vspace{0.1cm}\\
\emph{\textbf{Step $2.1$:} Proving the theorem for the last $K-1$ components:} \vspace{0.1cm}\\
Next, we show that the last $K-1$ components must be selected in decreasing order of $\pi_k(t_1)c_k/\mathbf{E}^{A2}(N_k|H_0)$ and tested by the SPRT. \\
Let
\beq
\gamma_j(t)=\frac{1}{\pi_j(t)\frac{f_j^{(1)}(\mathbf{y}_j(N_j))}{f_j^{(0)}(\mathbf{y}_j(N_j))}
          +1-\pi_j(t)}\;.
\eeq
Note that when the decision maker completes testing component $j$, the other components update their beliefs according to:
\beq
\pi_k(t_2)=\gamma_j(t_1)\pi_k(t_1) \;,\; \forall k\neq j \;.
\eeq
The expected cost achieved by $\boldsymbol\phi^{(j)}$ given the outcome (at time $t_2$) by testing component $j$ (i.e., given the observations vector $\mathbf{y}_j(N_j)$) is given by:
\beq
\label{eq:pr2_phi_prime_prime_exc1}
\bea{l}
\E\left\{\displaystyle\sum_{k=1}^{K}{c_k \tau_k \mathbf{1}_{\left\{k\in \mathcal{H}_1\right\}}}\;|\;\boldsymbol\phi^{(j)}, \mathbf{y}_j(N_j)\right\} \vspace{0.1cm}\\
=\pi_j(t_2)c_j N_j % \vspace{0.1cm}\\ \hspace{0.0cm}
+\left(1-\pi_j(t_2)\right)\times \vspace{0.1cm}\\ \hspace{0.3cm}
\E\left\{\displaystyle\sum_{k=1, k\neq j}^{K}{c_k \tau_k \mathbf{1}_{\left\{k\in \mathcal{H}_1\right\}}}\;|\;\boldsymbol\phi^{(j)}, \mathbf{y}_j(N_j), j\in \mathcal{H}_0\right\} \;.
\ena
\eeq
Let
\beq
\widetilde{\tau}_k=\tau_k-N_j \;\;\; \forall k\neq j
\eeq
be the modified stopping time, defined as the stopping time from $t=N_j+1$ until testing of component $k$ is completed. Thus, we can rewrite (\ref{eq:pr2_phi_prime_prime_exc1}) as:
\beq
\label{eq:pr2_phi_prime_prime_exc2}
\bea{l}
\E\left\{\displaystyle\sum_{k=1}^{K}{c_k \tau_k \mathbf{1}_{\left\{k\in \mathcal{H}_1\right\}}}\;|\;\boldsymbol\phi^{(j)}, \mathbf{y}_j(N_j)\right\} \vspace{0.1cm}\\
=\displaystyle\sum_{k=1}^{K}{\pi_k(t_2) c_k N_j} %\vspace{0.1cm}\\ \hspace{0.0cm}
+\left(1-\pi_j(t_2)\right)\times \vspace{0.1cm}\\ \hspace{0.3cm}
\E\left\{\displaystyle\sum_{k=1, k\neq j}^{K}{c_k \widetilde{\tau}_k \mathbf{1}_{\left\{k\in \mathcal{H}_1\right\}}}\;|\;\boldsymbol\phi^{(j)}, \mathbf{y}_j(N_j), j\in \mathcal{H}_0\right\} \;.
\ena
\eeq
The term $\sum_{k=1}^{K}{\pi_k(t_2) c_k N_j}$ in (\ref{eq:pr2_phi_prime_prime_exc2}) follows since,
\beq
\bea{l}
\Pr\left(k\in \mathcal{H}_1\;|\;\boldsymbol\phi^{(j)}, \mathbf{y}_j(N_j), j\in \mathcal{H}_0\right) \vspace{0.1cm}\\ \hspace{0.0cm}
=\displaystyle\frac{\Pr\left(k\in \mathcal{H}_1, j\in \mathcal{H}_0\;|\;\boldsymbol\phi^{(j)}, \mathbf{y}_j(N_j), \right)}
{\Pr\left(j\in \mathcal{H}_0\;|\;\boldsymbol\phi^{(j)}, \mathbf{y}_j(N_j), \right)} \vspace{0.1cm}\\ \hspace{0.0cm}
=\displaystyle\frac{\Pr\left(k\in \mathcal{H}_1\;|\;\boldsymbol\phi^{(j)}, \mathbf{y}_j(N_j), \right)}
{\Pr\left(j\in \mathcal{H}_0\;|\;\boldsymbol\phi^{(j)}, \mathbf{y}_j(N_j), \right)} =\displaystyle\frac{\pi_k(t_2)}
{1-\pi_j(t_2)} \triangleq \widetilde{\pi}_k(t_2)\;.
\ena
\eeq
Minimizing
\beq
\E\left\{\displaystyle\sum_{k=1}^{K}{c_k \tau_k \mathbf{1}_{\left\{k\in \mathcal{H}_1\right\}}}\;|\;\boldsymbol\phi^{(j)}, \mathbf{y}_j(N_j)\right\}
\eeq
at time $t_2$, requires one to minimize
\beq
\label{eq:modified_completion}
\E\left\{\displaystyle\sum_{k=1, k\neq j}^{K}{c_k \widetilde{\tau}_k \mathbf{1}_{\left\{k\in \mathcal{H}_1\right\}}}\;|\;\boldsymbol\phi^{(j)}, \mathbf{y}_j(N_j), j\in \mathcal{H}_0\right\}
\eeq
in (\ref{eq:pr2_phi_prime_prime_exc2}). \\
Note that (\ref{eq:modified_completion}) is the cost for $K-1$ components (where one and only one component is abnormal) starting at time $t=t_2=N_j+1$, with prior probability $\widetilde{\pi}_k(t_2)=\frac{\pi_k(t_2)}{1-\pi_j(t_2)}$ for component $k\neq j$ being abnormal. By the induction hypothesis, for any optimal selection rule $\boldsymbol\phi^{(j)}$ that selects component $j$ first, arranging the last $K-1$ components in decreasing order of $\widetilde{\pi}_k(t_2)c_k/\mathbf{E}^{A2}(N_k|H_0)$ (and testing them by the SPRT) minimizes (\ref{eq:modified_completion}). \\
Since
\beq
\widetilde{\pi}_k(t_2)=\frac{\gamma_j(t_1)}{1-\pi_j(t_2)}\pi_k(t_1) \;\;\forall k\neq j,
\eeq
then
\beq
\bea{l}
\displaystyle\frac{\widetilde{\pi}_1(t_2)c_1}{\mathbf{E}^{A2}(N_1|H_0)}\geq\frac{\widetilde{\pi}_2(t_2)c_2}{\mathbf{E}^{A2}(N_2|H_0)}\geq \cdots \geq\displaystyle\frac{\widetilde{\pi}_{j-1}(t_2)c_{j-1}}{\mathbf{E}^{A2}(N_{j-1}|H_0)} \vspace{0.1cm} \\
\geq\displaystyle\frac{\widetilde{\pi}_{j+1}(t_2)c_{j+1}}{\mathbf{E}^{A2}(N_{j+1}|H_0)}\geq\cdots
\geq\frac{\widetilde{\pi}_K(t_2)c_K}{\mathbf{E}^{A2}(N_K|H_0)}. \\
\ena
\eeq
Thus, the last $K-1$ components must be selected in decreasing order of $\pi_k(t_1)c_k/\mathbf{E}^{A2}(N_k|H_0)$ and tested by the SPRT.
\vspace{0.1cm}\\
\emph{\textbf{Step $2.2$:} Proving the theorem for all the $K$ components:} \vspace{0.1cm}\\
Finally, we show that component $1$ (i.e., the component with the highest index) must be selected first. The expected cost achieved by $(\boldsymbol\tau'(t), \boldsymbol\delta'(t), \boldsymbol\phi^{(j)})$ is given by:
\beq
\label{eq:pr2_phi_prime_prime_exc}
\bea{l}
\E\left\{\displaystyle\sum_{k=1}^{K}{c_k \tau_k \mathbf{1}_{\left\{k\in \mathcal{H}_1\right\}}}\;|\;(\boldsymbol\tau'(t), \boldsymbol\delta'(t), \boldsymbol\phi^{(j)})\right\} \vspace{0.1cm}\\
=\pi_j(t_1)c_j\left(\mathbf{E}'(N_j|H_1, t_1)\right) %\vspace{0.1cm}\\ \hspace{0.0cm}
+\displaystyle\sum_{k=1, k\neq j}^{K}\left[\pi_k(t_1)c_k \times \right. \vspace{0.1cm}\\ \left. \hspace{0.0cm}
\left(\mathbf{E}'\left(N_j | H_0, t_1\right)+\left(\displaystyle\sum_{i=1, i\neq j}^{k-1}{\mathbf{E}^{A2}\left(N_i|H_0\right)}\right)
 \right.\right. \vspace{0.1cm} \\ \left. \left. \hspace{5cm}
+\mathbf{E}^{A2}\left(N_k|H_1\right)\right)\right]\;.
\ena
\eeq
First, note that the expected cost achieved by $(\boldsymbol\tau'(t), \boldsymbol\delta'(t), \boldsymbol\phi^{(j)})$ can be further reduced for all $j$ by minimizing the expected sample size $\mathbf{E}'(N_j|H_i, t_1)$ for $i=0,1$,
which is achieved by $(\tau^{A2}_j, \delta^{A2}_j)$. Therefore, an optimal solution must be $(\boldsymbol\tau^{A2}, \boldsymbol\delta^{A2}, \boldsymbol\phi^{(j)})$ for an optimal selection rule $\boldsymbol\phi^{(j)}$. Thus, in the following we consider solutions of the form $(\boldsymbol\tau^{A2}, \boldsymbol\delta^{A2}, \boldsymbol\phi)$. \\
Next, by contradiction, consider an optimal selection rule $\boldsymbol\phi^{(j\neq 1)}$ that selects component $j\neq 1$ first. Therefore, $\boldsymbol\phi^{(j\neq 1)}$ selects the components in the following order:
\begin{center}
$j, 1, 2, ..., j-1, j+1, ..., K$.
\end{center}
As a result, the expected cost achieved by $(\boldsymbol\tau^{A2}, \boldsymbol\delta^{A2}, \boldsymbol\phi^{(j\neq 1)})$ is given by:
\beq
\label{eq:pr2_phi_exc}
\bea{l}
\E\left\{\displaystyle\sum_{k=1}^{K}{c_k \tau_k \mathbf{1}_{\left\{k\in \mathcal{H}_1\right\}}}\;|\;(\boldsymbol\tau^{A2}, \boldsymbol\delta^{A2}, \boldsymbol\phi^{(j\neq 1)})\right\} \vspace{0.1cm}\\
=\pi_j(t_1)c_j\left(\mathbf{E}^{A2}(N_j|H_1)\right) \vspace{0.1cm}\\ \hspace{0.5cm}
+\pi_1(t_1)c_1\left[\mathbf{E}^{A2}\left(N_j|H_0\right)+\mathbf{E}^{A2}\left(N_1|H_1\right)\right]
\vspace{0.1cm}\\ \hspace{0.5cm}
+\displaystyle\sum_{k=2, k\neq j}^{K}\left[\pi_k(t_1)c_k \times \right. \vspace{0.1cm}\\ \left. \hspace{0.5cm}
\left(\mathbf{E}^{A2}\left(N_j|H_0\right)+\left(\displaystyle\sum_{i=1, i\neq j}^{k-1}{\mathbf{E}^{A2}\left(N_i|H_0\right)}\right)
 \right.\right. \vspace{0.1cm} \\ \left. \left. \hspace{5cm}
+\mathbf{E}^{A2}\left(N_k|H_1\right)\right)\right]\;.
\ena
\eeq
We use the interchange argument to prove the theorem. Consider a selection rule $\boldsymbol\phi^{(1)}$ that selects component $1$ first followed by components $j, 2, 3, j-1, j+1, ..., K$. Similar to (\ref{eq:pr2_phi_exc}), the expected cost achieved by $(\boldsymbol\tau^{A2}, \boldsymbol\delta^{A2}, \boldsymbol\phi^{(1)})$ is given by:
\beq
\label{eq:pr2_phi_prime_exc}
\bea{l}
\E\left\{\displaystyle\sum_{k=1}^{K}{c_k \tau_k \mathbf{1}_{\left\{k\in \mathcal{H}_1\right\}}}\;|\;(\boldsymbol\tau^{A2}, \boldsymbol\delta^{A2}, \boldsymbol\phi^{(1)})\right\} \vspace{0.1cm}\\
=\pi_1(t_1)c_1\left(\mathbf{E}^{A2}(N_1|H_1)\right) \vspace{0.1cm}\\ \hspace{0.5cm}
+\pi_j(t_1)c_j\left[\mathbf{E}^{A2}\left(N_1|H_0\right)+\mathbf{E}^{A2}\left(N_j|H_1\right)\right]
\vspace{0.1cm}\\ \hspace{0.5cm}
+\displaystyle\sum_{k=2, k\neq j}^{K}\left[\pi_k(t_1)c_k \times \right. \vspace{0.1cm}\\ \left. \hspace{0.5cm}
\left(\mathbf{E}^{A2}\left(N_j|H_0\right)+\left(\displaystyle\sum_{i=1, i\neq j}^{k-1}{\mathbf{E}^{A2}\left(N_i|H_0\right)}\right)
 \right.\right. \vspace{0.1cm} \\ \left. \left. \hspace{5cm}
+\mathbf{E}^{A2}\left(N_k|H_1\right)\right)\right]\;.
\ena
\eeq

By comparing (\ref{eq:pr2_phi_exc}) and (\ref{eq:pr2_phi_prime_exc}), it can be verified that:
\begin{center}
$\bea{l}
\E\left\{\displaystyle\sum_{k=1}^{K}{c_k \tau_k \mathbf{1}_{\left\{k\in \mathcal{H}_1\right\}}}\;|\;(\boldsymbol\tau^{A2}, \boldsymbol\delta^{A2}, \boldsymbol\phi^{(1)})\right\} \vspace{0.1cm}\\
\hspace{2cm}
\leq \E\left\{\displaystyle\sum_{k=1}^{K}{c_k \tau_k \mathbf{1}_{\left\{k\in \mathcal{H}_1\right\}}}\;|\;(\boldsymbol\tau^{A2}, \boldsymbol\delta^{A2}, \boldsymbol\phi^{(j\neq 1)})\right\}
\ena$
\end{center}
since
$\pi_1(t_1)c_1/\mathbf{E}^{A2}(N_1|H_0)\geq\pi_j(t_1)c_j/\mathbf{E}^{A2}(N_j|H_0)$\;. \\
The expected cost can be reduced by selecting component $1$ first followed by component $j$, which contradicts the optimality of $\boldsymbol\phi^{(j\neq 1)}$. Hence, at time $t_1$ selecting component $1$ minimizes the expected cost. We have already proved that selecting the last $K-1$ components in decreasing order of $\pi_k(t_1)c_k/\mathbf{E}^{A2}(N_k|H_0)$ minimizes the objective function, which completes the proof.
\newcommand*{\QEDA}{\hfill\ensuremath{\blacksquare}}%
\QEDA

\subsection{Proof of Theorem \ref{th:optimality_alg1_2} Under The Independent Model}
\label{app:independent}
Let $\mathbf{E}'(N_k|H_i,$t$)$ be the expected sample size achieved by a stopping rule and a decision rule $(\tau'_k(t), \delta'_k(t))$, depending on the time that component $k$ is tested (i.e., $(\tau'_k(t), \delta'_k(t))$ depend on the selection rule), such that error constraints are satisfied. Let $\mathbf{E}^{A1}(N_k|H_i)$ be the expected sample size achieved by the SPRT's stopping rule and decision rule $(\tau^{A1}_k, \delta^{A1}_k)$, independent of the time that component $k$ is tested (i.e., $(\tau^{A1}_k, \delta^{A1}_k)$ are independent of the selection rule), such that error constraints are satisfied. Clearly, $\mathbf{E}^{A1}(N_k|H_i)\leq \mathbf{E}'(N_k|H_i,t)$ for all $k ,t$, for $i=0, 1$ and are achieved by the $\pi c N$-SPRT algorithm.

First, consider the case where $K=2$. Assume that
\begin{center}
$\displaystyle\frac{\pi_1(t_1)c_1}{\mathbf{E}^{A1}(N_1)}\geq\frac{\pi_2(t_1)c_2}{\mathbf{E}^{A1}(N_2)}$\;.
\end{center}
Consider selection rules $\boldsymbol\phi^{(1)}$, $\boldsymbol\phi^{(2)}$ that select component $1$ first followed by component $2$ and component $2$ first followed by component $1$, respectively. The expected cost achieved by $(\boldsymbol\tau'(t), \boldsymbol\delta'(t), \boldsymbol\phi^{(2)})$ is given by:
\beq
\label{eq:pr2_2_comp_t}
\bea{l}
\E\left\{\displaystyle\sum_{k=1}^{K}{c_k \tau_k \mathbf{1}_{\left\{k\in \mathcal{H}_1\right\}}}\;|\;(\boldsymbol\tau'(t), \boldsymbol\delta'(t), \boldsymbol\phi^{(2)})\right\} \vspace{0.1cm}\\ \hspace{0.5cm}
=\left(\mathbf{E}'(N_2|H_1, t_1)\right)\pi_2(t_1)c_2 \vspace{0.1cm}\\ \hspace{1cm}
+ \left(\mathbf{E}'(N_2|t_1) + \mathbf{E}'(N_1|H_1, t_2)\right)\pi_1(t_1) c_1.
\ena
\eeq
The expected cost achieved by $(\boldsymbol\tau'(t), \boldsymbol\delta'(t), \boldsymbol\phi^{(1)})$ is given by:
\beq
\bea{l}
\E\left\{\displaystyle\sum_{k=1}^{K}{c_k \tau_k \mathbf{1}_{\left\{k\in \mathcal{H}_1\right\}}}\;|\;(\boldsymbol\tau'(t), \boldsymbol\delta'(t), \boldsymbol\phi^{(1)})\right\} \vspace{0.1cm}\\ \hspace{0.5cm}
=\left(\mathbf{E}'(N_1|H_1, t_1)\right)\pi_1(t_1)c_1 \vspace{0.1cm}\\ \hspace{1cm}
+ \left(\mathbf{E}'(N_1 | t_1) + \mathbf{E}'(N_2|H_1, t_2)\right)\pi_2(t_1) c_2.
\ena
\eeq
Note that the expected cost achieved by both selection rules can be further reduced by minimizing the expected sample sizes (such that error constraints are satisfied) independent of the selection rules, which is achieved by $(\tau^{A1}_k, \delta^{A1}_k)$. Therefore, an optimal solution must be $(\boldsymbol\tau^{A1}, \boldsymbol\delta^{A1}, \boldsymbol\phi^{(1)})$ or $(\boldsymbol\tau^{A1}, \boldsymbol\delta^{A1}, \boldsymbol\phi^{(2)})$.
Next, we use the interchange argument to prove the theorem for $K=2$.
The expected cost achieved by $(\boldsymbol\tau^{A1}, \boldsymbol\delta^{A1}, \boldsymbol\phi^{(2)})$ is given by:
\beq
\bea{l}
\E\left\{\displaystyle\sum_{k=1}^{K}{c_k \tau_k \mathbf{1}_{\left\{k\in \mathcal{H}_1\right\}}}\;|\;(\boldsymbol\tau^{A1}, \boldsymbol\delta^{A1}, \boldsymbol\phi^{(2)})\right\} \vspace{0.1cm}\\ \hspace{0.5cm}
=\left(\mathbf{E}^{A1}(N_2|H_1)\right)\pi_2(t_1)c_2 \vspace{0.1cm}\\ \hspace{1cm}
+ \left(\mathbf{E}^{A1}(N_2) + \mathbf{E}^{A1}(N_1|H_1)\right)\pi_1(t_1) c_1.
\ena
\eeq
The expected cost achieved by $(\boldsymbol\tau^{A1}, \boldsymbol\delta^{A1}, \boldsymbol\phi^{(1)})$ is given by:
\beq
\label{eq:pr2_2_comp}
\bea{l}
\E\left\{\displaystyle\sum_{k=1}^{K}{c_k \tau_k \mathbf{1}_{\left\{k\in \mathcal{H}_1\right\}}}\;|\;(\boldsymbol\tau^{A1}, \boldsymbol\delta^{A1}, \boldsymbol\phi^{(1)})\right\} \vspace{0.1cm}\\ \hspace{0.5cm}
=\left(\mathbf{E}^{A1}(N_1|H_1)\right)\pi_1(t_1)c_1 \vspace{0.1cm}\\ \hspace{1cm}
+ \left(\mathbf{E}^{A1}(N_1) + \mathbf{E}^{A1}(N_2|H_1)\right)\pi_2(t_1) c_2.
\ena
\eeq
The expected cost achieved by $\boldsymbol\phi^{(1)}$ is lower than that achieved by $\boldsymbol\phi^{(2)}$ since $\frac{\pi_1(t_1)c_1}{\mathbf{E}^{A1}(N_1)}\geq\frac{\pi_2(t_1)c_2}{\mathbf{E}^{A1}(N_2)}$, which completes the proof for $K=2$.\\
The rest of the proof follows by induction on the number of components, as was done under the exclusive model.
\QEDA

\subsection{Proof of Theorem \ref{th:asymptotic3}}
\label{app:comp_ind}
For every $k$, let $\mathbf{E}^*(N_k|H_i)$ be the minimal expected sample size that can be achieved by any sequential test, such that error constraints are satisfied.
Let $\mathbf{E}^{A3}(N_k|H_i)$ be the expected sample size achieved by the $\pi c N$-SGLRT/SALRT algorithm, such that error constraints are satisfied.
Clearly, $\mathbf{E}^*(N_k|H_i)\leq \mathbf{E}^{A3}(N_k|H_i)$ for all $k$, for $i=0, 1$. \\
Assume that
\beq
\label{eq:th5_order}
\displaystyle\frac{\pi_1(t_1)c_1}{\mathbf{E}^*(N_1)}\geq\frac{\pi_2(t_1)c_2}{\mathbf{E}^*(N_2)}\geq ... \geq\frac{\pi_K(t_1)c_K}{\mathbf{E}^*(N_K)}.
\eeq
Similar to the proof of Theorem \ref{th:optimality_alg1_2}, it can be verified that the optimal solution to (\ref{eq:opt1}) is to select the components in the following order: $1, 2, ..., K$, where the components are tested by a sequential test that achieves expected sample size $\mathbf{E}^*(N_k|H_i)$ for all $k$, for $i=0, 1$.
Therefore, the expected cost achieved by $(\boldsymbol\tau^*, \boldsymbol\delta^*, \boldsymbol\phi^*)$ is given by:
\beq
\label{eq:pr5_star}
\bea{l}
\E\left\{\displaystyle\sum_{k=1}^{K}{c_k \tau_k \mathbf{1}_{\left\{k\in \mathcal{H}_1\right\}}}\;|\;(\boldsymbol\tau^*, \boldsymbol\delta^*, \boldsymbol\phi^*)\right\} \vspace{0.1cm}\\
=
\displaystyle\sum_{k=1}^{K}\pi_k(t_1)c_k
\left[\left(\displaystyle\sum_{i=1}^{k-1}{\mathbf{E}^*\left(N_i\right)}\right)
+\mathbf{E}^*\left(N_k|H_1\right)\right].
\ena
\eeq
By the asymptotic optimality property of the SALRT/SGLRT for a single process (used in the $\pi c N$-SGLRT/SALRT algorithm), it follows that $\mathbf{E}^{A3}(N_k|H_i)\sim \mathbf{E}^*(N_k|H_i)$ for all $k$, for $i=0, 1$ as $P_k^{FA}\rightarrow 0, P_k^{MD}\rightarrow 0$.
As a result, for sufficiently small error probabilities, the solution $(\boldsymbol\tau^{A3}, \boldsymbol\delta^{A3}, \boldsymbol\phi^{A3})$ is to select the components in the following order: $1, 2, ..., K$, where the components are tested by an asymptotically optimal sequential test that achieves expected sample size $\mathbf{E}^{A3}(N_k|H_i)$ for all $k$, for $i=0, 1$.
Therefore, the expected cost achieved by $(\boldsymbol\tau^{A3}, \boldsymbol\delta^{A3}, \boldsymbol\phi^{A3})$ is given by:
\beq
\label{eq:pr5_prime}
\bea{l}
\E\left\{\displaystyle\sum_{k=1}^{K}{c_k \tau_k \mathbf{1}_{\left\{k\in \mathcal{H}_1\right\}}}\;|\;(\boldsymbol\tau^{A3}, \boldsymbol\delta^{A3}, \boldsymbol\phi^{A3})\right\} \vspace{0.1cm}\\
=
\displaystyle\sum_{k=1}^{K}\pi_k(t_1)c_k
\left[\left(\displaystyle\sum_{i=1}^{k-1}{\mathbf{E}^{A3}\left(N_i\right)}\right)
+\mathbf{E}^{A3}\left(N_k|H_1\right)\right].
\ena
\eeq
Since $\mathbf{E}^{A3}(N_k|H_i)\sim \mathbf{E}^*(N_k|H_i)$ for $i=0, 1$ as $P_k^{FA}\rightarrow 0, P_k^{MD}\rightarrow 0$ for all $k$, the theorem follows.
\QEDA

\subsection{Proof of Theorem \ref{th:asymptotic4}}
\label{app:comp_exc}
The structure of the proof is similar to the proof of Theorem \ref{th:asymptotic3}. Hence, we provide a sketch of the proof, using notation similar to that used in the proof of Theorem \ref{th:asymptotic3}. Similar to the proof of Theorem \ref{th:optimality_alg1_2}, it can be verified that the optimal solution to (\ref{eq:opt1}) is to select the components in decreasing order of $\pi_k(t_1)c_k/\mathbf{E}^*(N_k|H_0)$, where the components are tested by a sequential test that achieves expected sample size $\mathbf{E}^*(N_k|H_i)$ for all $k$, for $i = 0, 1$. By the asymptotic optimality property for a single process of the SALRT/SGLRT (used in the $\pi c N_0$-SGLRT/SALRT algorithm), it follows that $\mathbf{E}^{A4}(N_k|H_i)\sim \mathbf{E}^*(N_k|H_i)$ for all $k$, for $i=0, 1$ as $P_k^{FA}\rightarrow 0, P_k^{MD}\rightarrow 0$. As a result, for sufficiently small error probabilities, the solution $(\boldsymbol\tau^{A4}, \boldsymbol\delta^{A4}, \boldsymbol\phi^{A4})$ is to select the components in decreasing order of $\pi_k(t_1)c_k/\mathbf{E}^*(N_k|H_0)$, where the components are tested by an asymptotically optimal sequential test that achieves expected sample size $\mathbf{E}^{A4}(N_k|H_i)$ for all $k$, for $i=0, 1$. Similar to the proof of Theorem \ref{th:asymptotic3}, comparing the objective functions achieved by $(\boldsymbol\tau^*, \boldsymbol\delta^*, \boldsymbol\phi^*)$ and $(\boldsymbol\tau^{A4}, \boldsymbol\delta^{A4}, \boldsymbol\phi^{A4})$ proves the theorem.
\QEDA

\bibliographystyle{ieeetr}
%\bibliography{anomaly_detection_bib}

\end{document}